\documentclass[conference]{IEEEtran}
\IEEEoverridecommandlockouts
\usepackage{cite}
\usepackage{amsmath,amssymb,amsfonts}
\usepackage{algorithmic}
\usepackage{graphicx}
\usepackage{textcomp}
\usepackage{xcolor,siunitx}
\def\BibTeX{{\rm B\kern-.05em{\sc i\kern-.025em b}\kern-.08em
    T\kern-.1667em\lower.7ex\hbox{E}\kern-.125emX}}
\begin{document}

\title{Improving performance of logical qubits by parameter tuning and topology compensation}

\author{\IEEEauthorblockN{1\textsuperscript{st} Jack Raymond}
\IEEEauthorblockA{\textit{D-Wave Systems} \\
  Burnaby, Canada \\
  ORCID 0000-0002-1808-6039}
\and
\IEEEauthorblockN{2\textsuperscript{nd} Ndiam{\'e} Ndiaye}
\IEEEauthorblockA{\textit{McGill University} \\
  Montreal, Canada \\
  ORCID: 0000-0002-4920-6566}
\and
\IEEEauthorblockN{3\textsuperscript{rd} Gautam Rayaprolu}
\IEEEauthorblockA{\textit{McGill University} \\
  Montreal, Canada \\
  ORCID 0000-0001-8795-8614}
\and
\IEEEauthorblockN{4\textsuperscript{th} Andrew~D.~King}
\IEEEauthorblockA{\textit{D-Wave Systems} \\
  Burnaby, Canada \\
  ORCID 0000-0001-8362-8941}
}

\maketitle

\begin{abstract}
  Optimization or sampling of arbitrary pairwise Ising models, in a quantum annealing protocol of constrained interaction topology, can be enabled by a minor-embedding procedure.
  The logical problem of interest is transformed to a physical (device programmable) problem, where one binary variable is represented by a logical qubit consisting of multiple physical qubits.
  In this paper we discuss tuning of this transformation for the cases of clique, biclique, and cubic lattice problems on the D-Wave 2000Q quantum computer.
  We demonstrate parameter tuning protocols in spin glasses and channel communication problems, focusing on anneal duration, chain strength, and mapping from the result on physical qubits back to the logical space.
  Inhomogeneities in effective coupling strength arising from minor-embedding are shown to be mitigated by an efficient reweighting of programmed couplings, accounting for logical qubit topology.
\end{abstract}


\section{Introduction}

Quantum annealing can be applied for the purposes of optimization or sampling of complicated distributions~\cite{Kadowaki1998,Farhi2001}.
Quantum processing units (QPUs) executing this algorithm have recently been realized at scale in specialized hardware~\cite{Johnson2011}. 
Interest in quantum annealing stems from the potential for differentiated absolute or scaling performance advantage in practical problems, over algorithms implemented on classical hardware. This potential is subject to intense theoretical and empirical testing~\cite{McGeoch2013,Ronnow:18,Boixo2016,albash:18,King:19}. 
Optimization and sampling problems to which D-Wave computers have been applied include magnetic material simulations, quantum Boltzmann machine learning, traffic routing, multi-user channel decoding, and many others~\cite{Harris162,King2018,Vinci2019,Neukart2017,D-WaveSystems,10.1145/3341302.3342072}. 
While quantum annealing can in principle apply to a broad range of optimization problems, practical QPUs work within engineering constraints. Many interesting applications must be transformed for compatibility. 

The target problems we wish to optimize or sample by quantum annealing are described by an Ising Hamiltonian: a function over $N$ binary spin variables $x \in \{-1,1\}^N$
\begin{equation}
  H(x) = \sum_{a<b} J_{a,b} x_a x_b + \sum_a h_a x_a\label{eq:Hising}\;,
\end{equation}
where $J$ and $h$ are couplers and external fields respectively.
An optimization problem with respect to $H(x)$ is to determine $x^* = \mathrm{argmin}\{H(x)\}$. This problem is also commonly referred to as quadratic unconstrained binary optimization. 
Determining optima for this model, fair sampling at low energy, and many other inference problems are NP-hard~\cite{Barahona_1982,Jerrum1986};
many random problems have an energy landscape that is challenging to all known heuristics.
We consider in this paper two exemplars of these hard random optimization problems: spin glasses and code-division multiple access (CDMA)~\cite{Nishimori2001,TanakaCDMA}.

Quantum annealing solves the optimization problem by evolving a physical state over a total annealing time $t_a$, through a time-dependent Hamiltonian parameterized by $s=t/t_a$ given by 
\begin{equation}
  {\hat H}(s) \!=\! A(s)\left[ - \sum_a \sigma_a^x \right] + B(s)\!\left[\sum_{a<b} J_{a,b} \sigma^z_a \sigma^z_b + \sum_a h_a \sigma^z_a \right] \label{Hising2}\;,
\end{equation}
where $A(s)$ and $B(s)$ are transverse and longitudinal energy scales respectively, and $\sigma^{x,z}$ are Pauli matrices with $z$ signifying the computational basis.
By slow physical evolution from the prepared ground state at $s=0$, with $A(s\!=\!0)\gg B(s\!=\!0)$, to $A(s\!=\!1) \ll B(s\!=\!1)$, the state is guaranteed to concentrate over ground states of (\ref{eq:Hising}), so that a measurement at $s=1$ yields with high probability a ground state.
While asymptotic guarantees exist for error-free annealing at low temperature, restarting the annealing process multiple times (with shorter durations) is normally advantageous for minimizing time-to-solution. In practice, suboptima are frequently encountered, and quantum annealing provides a distribution of states not necessarily dominated by optima.

In this paper we study problems on a QPU that are transformed by a minor-embedding process \cite{Choi2008,Choi2011}. With consideration of anneal duration, chain strength and minor-embedding refinement we show that spin-glass and CDMA problems can be solved in typical cases at the largest programmable size in a single programming cycle.
We demonstrate how logical qubit topology leads to asymmetric effective coupling strengths, but can be compensated for by simple heuristics.

In Sections \ref{sec:SGCDMA} and \ref{sec:minorembedding} we describe the problems considered and their minor-embedding. In Sections \ref{sec:cs}, \ref{sec:ptlsm}, and \ref{sec:oad} we discuss optimization of chain strength, the mapping from physical to variable spaces, and anneal duration, demonstrating impact on our problem classes. Our main new result is presented in Section \ref{sec:ljpc}, where we predict and compensate for asymmetries in coupling brought about by the minor-embedding process. We conclude in Section \ref{sec:conclusion}.

\subsection{Spin Glasses and Code-Division Multiple Access}
\label{sec:SGCDMA}
In this paper we consider two paradigmatic random problems that, with appropriate parameters, are hard to optimize in typical cases.
Spin glasses have long been studied as exemplars of challenging energy landscapes, including in D-Wave QPUs~\cite{McGeoch2013,SKPRB,Venturelli2015a,Harris162}.
We consider clique (CSG), biclique (BSG) and 3D cubic lattice (3DSG) spin glasses. For all these models external fields are zero, $h_i=0$ $\forall i$. The CSG problem is defined by random couplings, $J_{ij} = \pm 1$ with equal probability. The BSG problem has its variables divided into two equal-sized sets, all couplings within the set are zero, and between sets are $\pm 1$ with equal probability. In the 3DSG problem non-zero couplings are restricted to edges of a cubic lattice with open boundary conditions.
We can define a challenging target energy $E_T$ for an instance of any of these models as an upper bound on the ground state energy, determined in this paper as the minimum energy found by any solver.

CDMA is the second model considered, where the optimization problem is to determine the most probable transmitted signal over a multi-user channel.
Given $N$ uniformly distributed binary variables ($b\in\{-1,1\}^N$) to be transmitted, an $M\times N$ code matrix of uniformly distributed binary variables ($W \in \{-\frac{1}{\sqrt{N}},\frac{1}{\sqrt{N}}\}^{M\times N}$), and a noise vector $n$ of $M$ normally distributed variables $n_i \sim N(0,1)$ we can define our instance. The received signal is constructed as $y= W b + \sigma_0 n$, where $1/2\sigma_0^2$ is the signal-to-noise ratio of the channel. The negative log-likelihood of the transmitted variables given the signal can be written $H(x) = \frac{1}{2\sigma_0^2}\sum_\mu \left(y_\mu - \sum_i W_{\mu i} x_i\right)^2$, which by expansion gives (\ref{eq:Hising}) (up to an irrelevant constant). At a load $M/N=1.4$, and a signal-to-noise ratio of $\SI{7}{dB}$, we operate close to a first-order phase transition point, where recovery of the transmitted bit sequence is challenging~\cite{TanakaCDMA}. As with the spin glass, the energy landscape is analytically characterized in the large $N$ limit, and closely related multi-user communication channel problems have been studied in the context of QPUs~\cite{10.1145/3341302.3342072}.
For an instance of CDMA we can define a target energy as $E_T = H(b)$ (the energy associated to the transmitted bit sequence). 

The main objective considered in this paper is to achieve the target energy. A secondary (and correlated) objective is removing asymmetries in the distribution of samples arising from minor-embedding.

\subsection{Minor-Embedding of Hamiltonians}
\label{sec:minorembedding}
D-Wave computers allow annealing routines of superconducting flux qubits, which can be modelled as Ising spins~\cite{Johnson2011}.
However, only a subset of edges is available for programming in the D-Wave 2000Q computer as described by a Chimera graph topology~\cite{Bunyk2014architectural}. As such we cannot code some of the interactions necessary in the target spin-glass and CDMA problems using one qubit per variable.
To solve the problem by annealing with the constrained topology we can employ minor-embedding~\cite{Choi2008,Choi2011}.
A minor-embedding is described by a set of logical qubits (one per variable in the target problem), with each logical qubit ($a$) described by a set of connected physical qubits $\mathcal{C}_a, a = 1,\ldots,N$.
Logical qubits meet the requirement that for any non-zero coupling in the target problem ($J_{ab}$) there exists at least one programmable coupler between the logical qubits: $\sum_{i \in S_a,j\in S_b} A_{i,j} > 0,\; \forall J_{ab} \neq 0$. $A$ is the QPU adjacency matrix ($A_{ij}=1$ if physical qubits $i$, $j$ are connected by a programmable coupler, $0$ otherwise).
Methods for generating efficient regular embeddings of clique and biclique problems are encoded as part of the D-Wave API~\cite{Ocean,cliquepaper}, which are employed here. For the cubic lattice we use a minor-embedding with 4 physical qubits per logical qubit~\cite{Harris162}.

The programmable topology of the online system DW\_2000Q\_6 is a C16 Chimera graph, meaning 16 by 16 cells each of 8 qubits, subject to a small number of defects~\cite{DW_2000Q_6}. We also consider in this paper problems programmed over a C8 (8 by 8 cell) subgraph with no defects.

\begin{figure}[htbp]
\centerline{\includegraphics[width=0.5\linewidth]{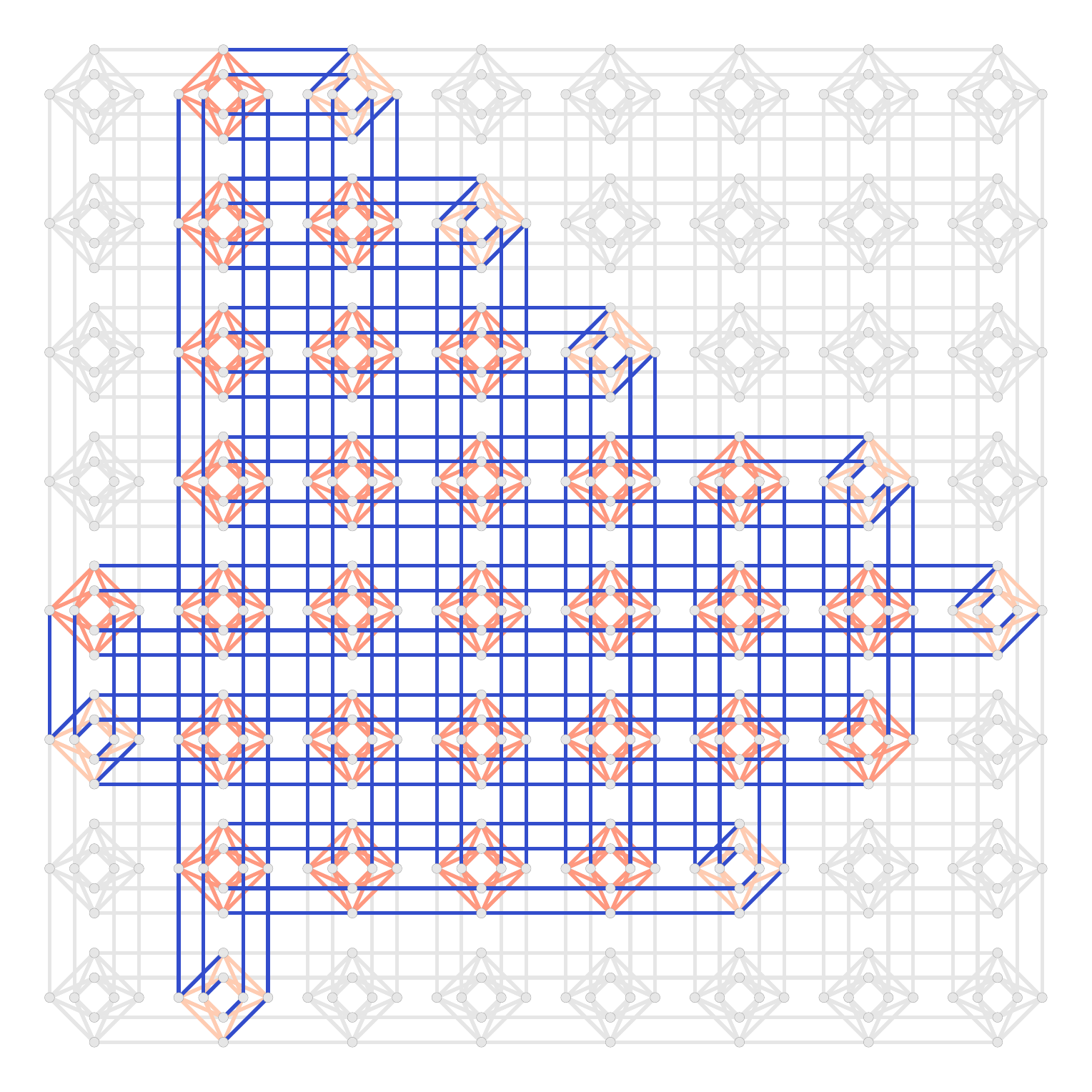}\includegraphics[width=0.5\linewidth]{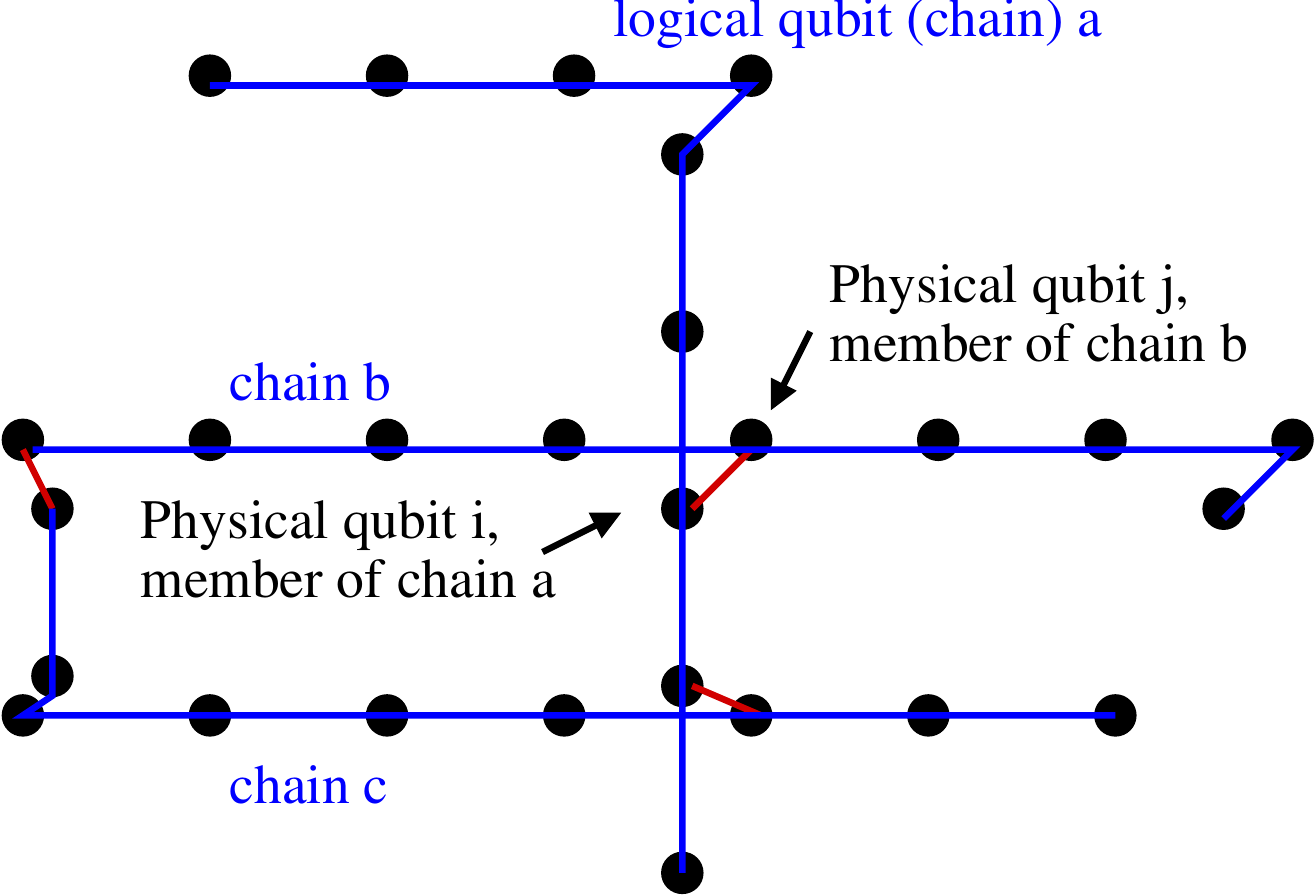}}
\caption{(left) A typical example of a $N=32$ variable embedding produced by find\_clique\_embedding Ocean tool over a C8 subgraph of DW\_2000Q\_6~\cite{cliquepaper}. Blue couplers are ferromagnetic and encode the logical qubits, orange couplers encode the target problem interactions among the logical qubits. (right) Consider a subset of 3 chains; chains $a$ and $b$ connect through physical qubits $i$ and $j$ that are relatively central. Chains $b$ and $c$ connect through peripheral qubits and have a weaker effective coupling when considering typical pairs of qubit on each chain.}
\label{fig:c8clique}
\end{figure}

Since each logical qubit in these embeddings consists of sequentially coupled qubits, we will refer to them as {\em chains}. The number of physical qubits comprising each chain is uniform for a given size and target topology, and this number will be referred to as the {\em chain length} ($L$). For each of the problem types considered, at fixed size $N$, the chain length is constant.
An example of a clique minor-embedding is shown in Fig.~\ref{fig:c8clique}. 
A clique of size $N$ requires chain length $L=\lceil N/4\rceil + 1$.

Given the chains, one can define the minor-embedded problem as
\begin{multline}
  H(z)  = R \left\lbrace\lambda \left[-\sum_{a}\sum_{i<j \in C_a} A_{ij} z_i z_j \right] + \right. \\\left. \left[\sum_{a,b} {\hat J}_{a,b} \sum_{i \in C_a,j \in C_b} A_{ij} z_i z_j + \sum_a {\hat h}_a \sum_{i \in C_a}  z_i\right]\right\rbrace\;.
\end{multline}
$R$ is a rescaling term necessary to ensure all Hamiltonian terms are within device programming ranges, $\lambda$ is a chain strength parameter, and ${\hat h}_a$, ${\hat J}_{ab}$ are programmed fields and couplings. The first term in the physical Hamiltonian promotes (for $\lambda>0$) alignment of physical qubits within a chain; when qubits align we can map the physical state to variables in the target problem as $x_a=\frac{1}{|C_a|}\sum_{i \in C_a} z_i = \pm 1$. Using a {\em uniform spreading} approach, we can in addition choose:
\begin{equation}
  {\hat h}_{a} = \frac{h_a}{|C_a|}\;, \qquad
  {\hat J}_{ab} = \frac{J_{ab}}{\sum_{i \in C_a,j \in C_b} A_{ij}}\;. \label{eq:us}
\end{equation}
This ensures there is a one-to-one mapping between chain-aligned states $z$, and target problem states of the same energy (up to an irrelevant constant offset). These choices ensure that the two spaces not only have the same ground states (provided $\lambda$ is large enough), but that any distribution over the physical space dependent only on energy will map to a similar distribution on the target space, restricting to the chain-aligned subspace. As an example, a Boltzmann distribution in the physical space begets a Boltzmann distribution in the logical space with the same temperature.

A finite programmable range exists for the QPU. In this study $R\lambda \leq 2$ exploiting extended {\em J}-range over chains, with a regular range restriction on other terms $R|{\hat J}_{ab}|\leq 1$, and $R|{\hat h}_{a}|\leq 2$. If chain breaks are penalized with large $\lambda$, then $R$ is scaled down, decreasing the problem energy scale relative to noise and temperature in the QPU. This increases susceptibility of low-energy chain-aligned states to diabatic transitions, thermal excitations and analog errors. An intermediate chain strength is generally preferable. Empirical investigation allows an optimal chain strength to be found, but common strategies also include use of a bound that ensures the ground state is chain-aligned (with the potential downside of scaling down the problem part excessively), or maximizing $R$ at $R\lambda=2$ (so that energy scales are maximized in both the chain and problem terms, with the potential downside that uninformative chain misaligned states energetically predominate)~\cite{King2014}.

Aside from the chain strength ($\lambda$), the anneal duration ($t_a$) and the selection of programmed values (${\hat h}$ and ${\hat J}$) we can consider more general mappings (or post-processing of output) from physical to logical states, including making use of chain-misaligned states.
Given that we will operate at finite anneal times, chain strengths, and temperatures, we should anticipate observing many suboptima. Chain misaligned states are exponentially more numerous than chain-aligned ones, and so will entropically dominate the solution space at all energies, except perhaps a narrow window near the ground state energy (at large $\lambda$).

A common practice is to transform all samples into aligned states by majority vote (MV), $x_a = \mathrm{sign}\left(\sum_{j \in C_a} z_j\right)$, breaking the $\mathrm{sign}(0)$ case randomly \cite{King2014,Vinci2015}. For these chain-misaligned states, it is not true that a physically low energy state will map to (similarly) low energy target states, but MV can significantly enhance the probability to see optimal solutions as later shown. It should be emphasized that the inclusion of voted chain-misaligned states produces a complicated distribution in general, and a Boltzmann distribution in the physical space will not beget one in the logical space.

\section{Parameter Tuning}

\subsection{Chain Strength}
\label{sec:cs}
CSG has been studied on a previous generation of D-Wave QPU~\cite{Venturelli2015a}, concluding that chain strength should be tuned as
\begin{equation}
  \lambda = \lambda_0  \sqrt{\sigma^2 N}\;,\label{eq:chainStrength}
\end{equation}
where $\sigma^2 = \frac{2}{N(N-1)}\sum_{a<b} J_{ab}^2$ is the variance of the coupling strength ($1$ for CSG). The motivation for this choice is linked to existence of a spin-glass phase transition, with optimal $\lambda_0$ determined empirically.

Beyond the phase transition argument the scaling of chain strength can also be motivated on the basis of local embedding topology, indicating the quantity given by (\ref{eq:chainStrength}) may be effective more generally at least as a heuristic: If the physical low-energy space is to align well with the logical low-energy space, most chains must consist of aligned physical states. In a frustrated problem like the spin glass different physical qubits within chains will be subject to random energy signals from their neighbors. The chain can be divided into two parts, so that each part receives a random signal from approximately equal number of neighboring chains. The central limit theorem dictates that the signal in each part will be zero mean, and variance approximately $N\sigma^2/2$. In combination these can create a random torque on the chain favouring misalignment. Thus for the central coupling to maintain the alignment of the two halves it must involve an energy penalty larger than the torque signal, and so scale as $\sqrt{N}\sigma$. To prevent breakages on most chains $\lambda_0\approx 1$ can suffice~\cite{Venturelli2015a}.

The phase transition and torque arguments generalize very well to the case of a biclique, where $\sigma^2 \approx 1/2$ (approximately half of couplings are 0, and half are $\pm 1$). In the case of a cubic lattice the central limit theorem argument does not apply as each logical qubit has connectivity at most $6$, but we can still take this rule as a crude heuristic with $\sigma^2 \approx 6/N$.

The CDMA problem we have chosen as our second example contrasts with the spin-glass models in many regards: It is subject to a random first-order phase transition as opposed to a second-order spin-glass transition, and has correlated non-zero fields and couplings. However, the marginal distribution for $J_{ij}$ is approximately zero-mean Gaussian distributed,\footnote{Distribution converges to a Gaussian with increasing $N$.} with variance $\sigma^2=N/[M \sigma_0^2]$.

We note that careful evaluation of coupling patterns between the chains, or tuning based on empirical outcomes, can improve over the choice (\ref{eq:chainStrength}), but as we will show this heuristic rule works well across all these models with $\lambda_0\approx 1$. 

\subsection{Physical-to-Logical Mapping}
\label{sec:ptlsm}

MV is a common choice to make use of chain misaligned samples, maintaining the interpretation of logical qubits for chain-aligned states. We consider in this paper three alternatives. The first is to retain only chain-aligned states from the sampling set (A). In the second we choose a random physical qubit from the chain as the logical state (R)---MV, R, and A all agree for chain-aligned states. The final form of mapping involves a local (energetic) relaxation in the logical space, we call this greedy descent (GD).

Greedy descent is valuable when we approach a target solution, but fail to achieve it owing to some local excitation or defective chain. A common scenario in practical annealing, particularly considering finite temperature, is to have isolated physical or logical qubit excitations around a ground state. A small number of physical qubit excitations on a chain is curable via majority vote, but an entire chain flip is not. Given samples in the logical space, greedy descent can be applied to both remove local logical excitations and correct misfunctioning chains. For a given sample we can first map into the logical space using MV,\footnote{Greedy descent improves R, A, and MV mappings. Only the combination with MV is demonstrated for brevity.} and can update variables as
\begin{equation}
  x_a = - \mathrm{sign}\left(\sum_b[J_{ab} + J_{ba}] x_b + h_a\right) \label{eq:GD}\;,
\end{equation}
breaking ties $\mathrm{sign}(0)$ to leave $x_a$ unchanged. We can say an update is applied if the sign of $x_a$ is changed. We can apply this rule in a fixed random sequence iteratively to every sample, until no more updates are effected.

Reading one physical qubit to represent the logical state (R) has the advantage of reduced read-out and minimal computation costs. MV involves reading all qubits and performing a simple parallelizable operation, potentially correcting some physical qubit errors in the process. Discarding misaligned states requires some computations comparable to MV, maintaining the more promising (and easier to interpret) samples. Greedy descent involves a mapping plus additional operations in the target problem space correcting a wider variety of local excitations. The scaling of these transformations from physical to logical space per sample are $O(N)$ for method R, and $O(N L)$ for MV, A and GD.\footnote{Chain length scales at best in proportion to logical connectivity over Chimera graphs. In strongly performing cases we anticipate only $O(1)$ GD updates need be applied per logical qubit. As a consequence scaling of GD in $N$ is at worst equal to MV, although practical implementation can be more complicated.} 

We demonstrate that MV does not add much value with respect to the simpler method R, but both these methods outperform A in reaching the target energy. We demonstrate that greedy descent significantly decreases energy, but efficiency of implementation may inform the most practical choice.

\subsection{Annealing Time}
\label{sec:oad}

The time required to draw $n$ samples from the QPU in our experiments is well described by
\begin{equation}
  t = \left[t_{p} + n(t_a + t_r + t_d)\right] + t_{m} + t_{n}\;. \label{eq:timings}
\end{equation}
The square-bracketed part is the QPU access time, which is the time charged to users. Parameters $t_{p}$, $t_r$ and $t_d$ are programming, read-out, and delay times respectively. Measurements indicate $t_d+t_r = \SI{219}{\micro s}$ for all experiments presented, and $t_{p} \approx \SI{10}{\milli s}$ fluctuates little between the experiments.\footnote{Since we consider protocols using only one programming per instance $t_p$ is independent of $n$.} The annealing time ($t_a$) is controlled as an input.
Parameter $t_{m}$ is the time used mapping from the physical states to the target variables (e.g., MV, GD - which in this study is done off-QPU); we exclude this from our analysis ($t_m=0$).
Other times ($t_n$) such as network latencies and queueing are also excluded from our analysis ($t_n=0$).  

Using either the number of samples at which the target sample is first observed ($n_{RT}$), or the frequency with which target achieving samples are seen ($p$), we can understand the efficiency of annealing as a function of anneal duration.
Given $n_{RT}$ we can directly evaluate a time-to-solution by (\ref{eq:timings}), taking $n=n_{RT}$. However, if samples are independent and identically distributed (IID) then $P(n_{RT}) = p (1-p)^{n_{RT}-1}$~\cite{King2014}, and we can leverage this for a lower-variance estimator.\footnote{Alternative estimators sensitive to the sample ordering indicate comparable results, within sampling error.} Exploiting the relationship between the two quantities for a given instance we can estimate samples to the target solution with confidence $X$ as~\cite{King2014,Venturelli2015a}
\begin{equation}
  {\hat n}_{RT}(X) = \frac{\log(1 - X)}{\log(1 - p)}. \label{eq:sts}
\end{equation} 
We can estimate $p$ as the empirical frequency of ground states based on a sample set drawn by a single programming.

The threshold $X=0.99$ is commonly used in empirical studies~\cite{King2014}. Note that the threshold choice $\log(1\!-\!X)$ has the effect of linearly rescaling the per-sample times, discounting constant offsets like programming time. Optimization of anneal duration to minimize time-to-solution is a trade-off between $p$, typically an increasing function of $t_a$ over the programmable annealing range, and $t$ which is a linear function of $t_a$. We find empirically for a variety of models studied that $p$ is a weakly increasing function of $t_a$. As a consequence a reasonable first guess to the optimal anneal time is given by
\begin{equation}
  t_a = t_r + t_d \label{eq:oad}\;.
\end{equation}
For $t_a \ll t_r + t_d$ there is effectively no time penalty to increasing $t_a$, whereas for $t_a \gg t_r + t_d$ the time penalty $t_a$ exceeds the gain in ground state rates $p(t_a)$. 


\subsection{Results}
We have argued that a good use of QPU resources can be achieved by choosing a chain strength (\ref{eq:chainStrength}) with $\lambda_0 \approx 1$, applying a greedy descent (GD) mapping (\ref{eq:GD}), and tuning anneal duration (\ref{eq:oad}). In this section we verify these heuristics, restricting attention to uniform spreading (\ref{eq:us}). In each experiment we present a median with respect to 100 random instances, using approximately 1 second of QPU sampling time ($\lfloor 10^6/(t_a + t_d + t_r) \rfloor$ samples, with timings measured in $\SI{}{\micro s}$). We have used exclusively the DW\_2000Q\_6 online system with default settings except where stated otherwise~\cite{DW_2000Q_6}. We use a fixed minor-embedding for all experiments of a given topology (clique, biclique or 3d cubic lattice) and size. Extended {\em J}-range ($R\lambda\leq 2$) is used over chain couplers, and outperformed regular range ($R\lambda\leq 1$). At near optimal chain strengths the bound is typically saturated $R\lambda=2$ (i.e., chain couplers are programmed as $-2$).

\begin{figure}[htbp]
  \centerline{\includegraphics[width=0.5\linewidth]{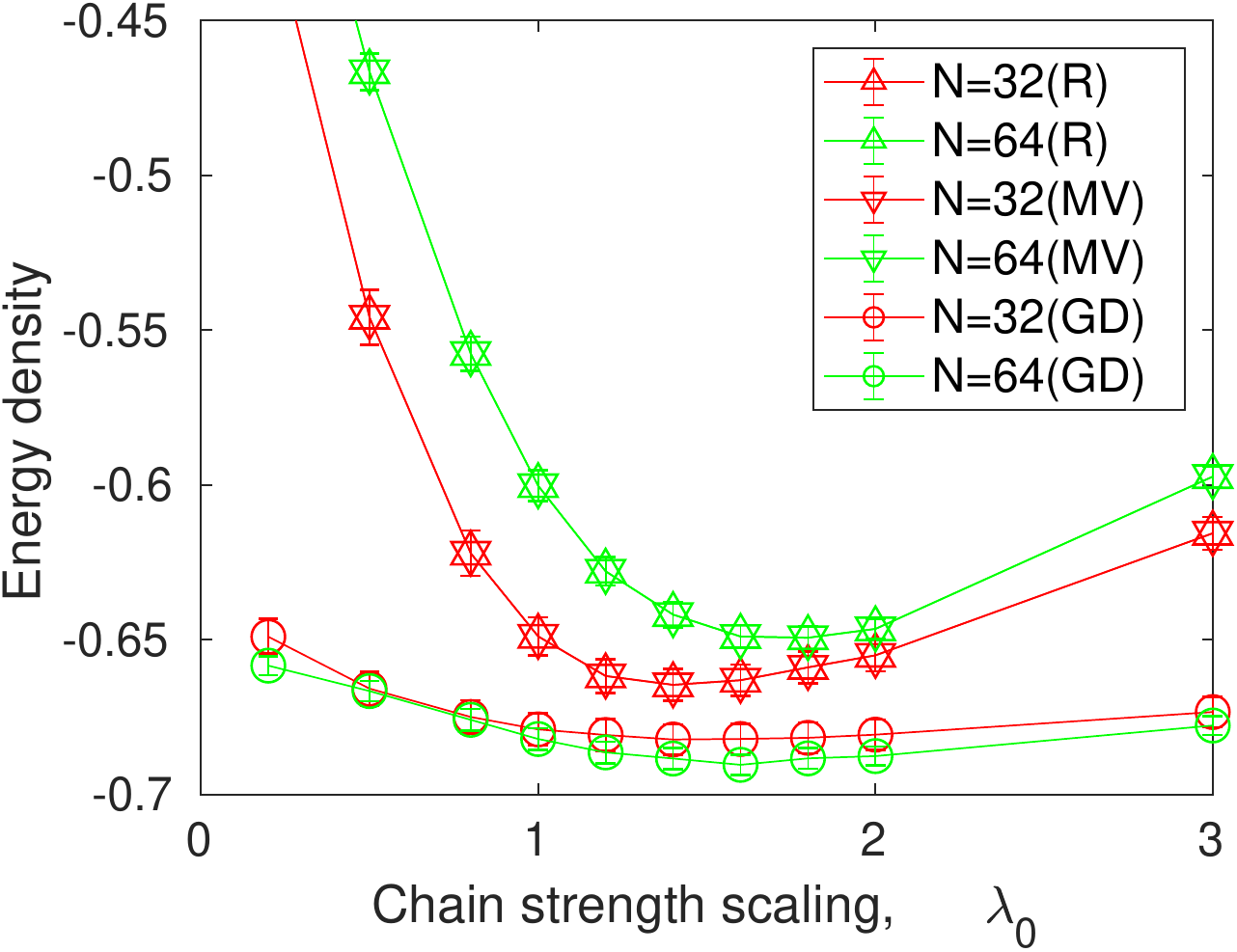}\includegraphics[width=0.5\linewidth]{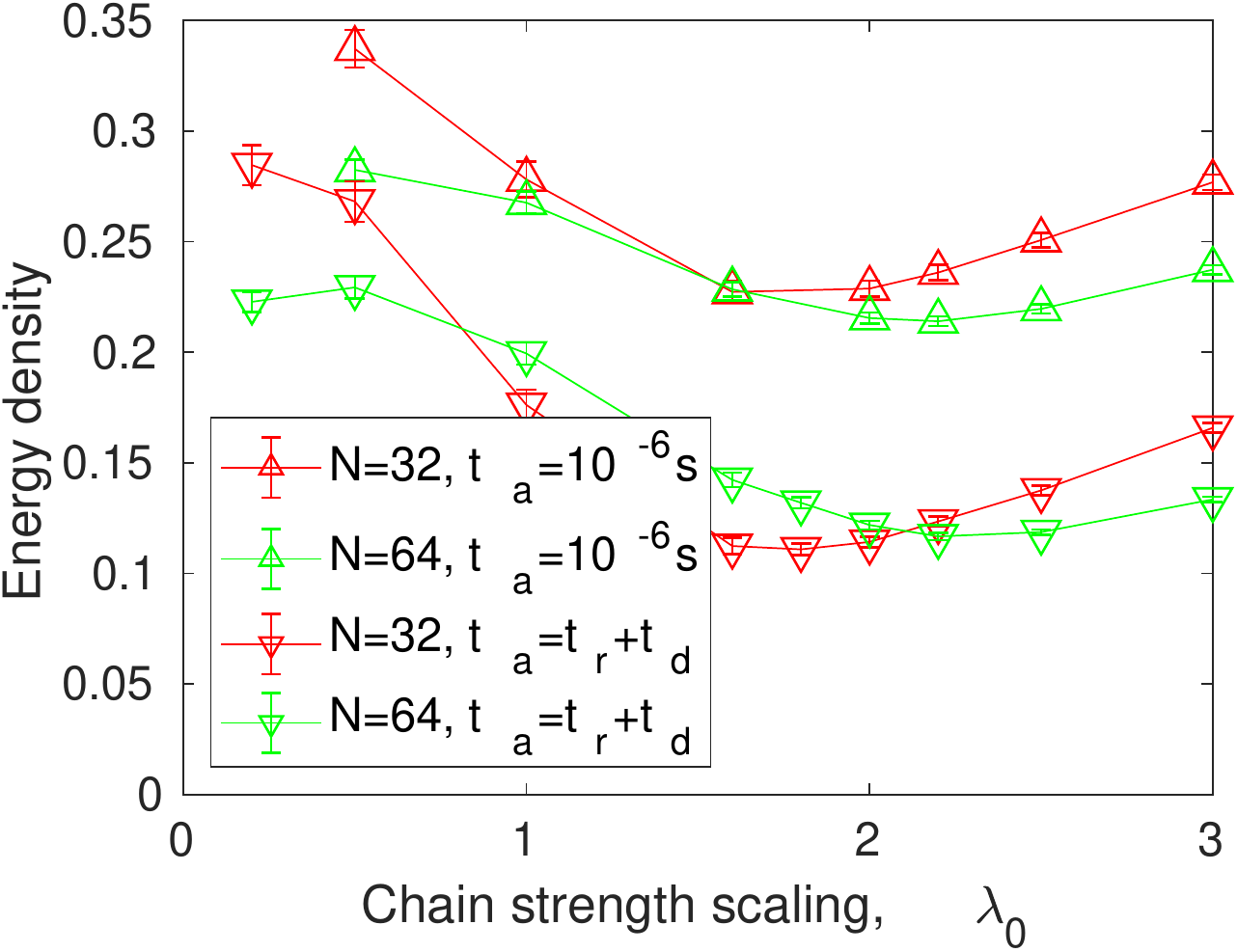}}
\caption{Conditions (\ref{eq:chainStrength}) predicts a reasonable chain strength with best performance in mean energy at $\lambda_0\gtrsim 1$ for CSG and CDMA models under a variety of parameters: (left) CSG at $t_a = t_d+t_r$, energy density $\langle H(x) \rangle N^{-3/2}$ (right) CDMA with MV samples only, energy density $[\langle H(x) \rangle-H(b)]N^{-3/2}$. In each case $\langle \rangle$ is the sample average and a normalization $N^{3/2}$ is applied to account for expected scaling.  Variability is lessened in the GD case as the mapping corrects errors, MV and R mappings are indistinguishable. Position of the optima shifts to larger values for larger sizes. }
\label{fig:chainStrength}
\end{figure}

In Fig.~\ref{fig:chainStrength} we present data on sample average energy as a function of chain strength. The curve minima indicate chain strengths minimizing the sample average energy of a typical (median) instance. The chain strength heuristic (\ref{eq:chainStrength}) works well.

\begin{figure}[htbp]
  \centerline{
    \includegraphics[width=0.5\linewidth]{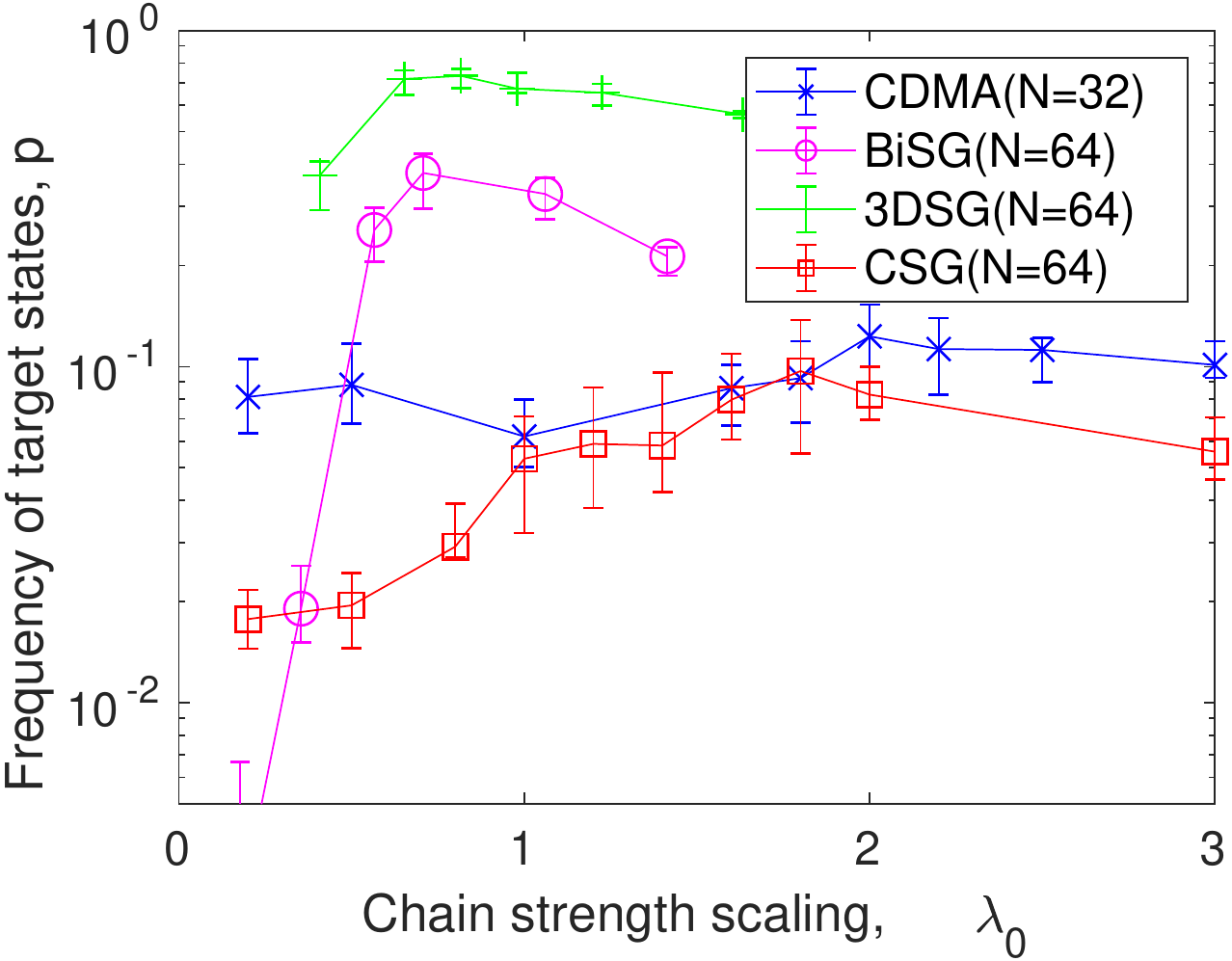}
    \includegraphics[width=0.5\linewidth]{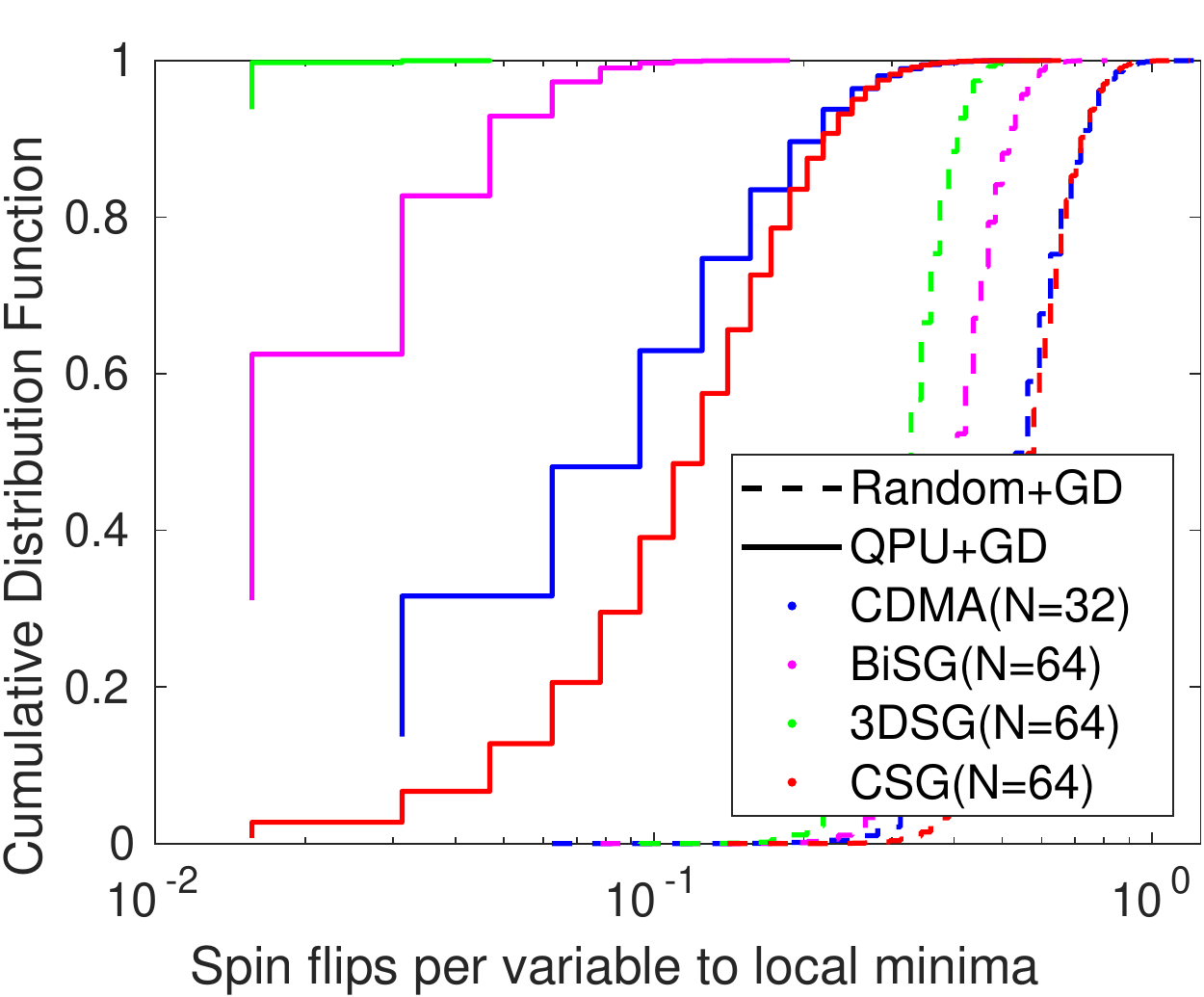}
  }
\caption{(left) We show $p$ versus chain strength with GD mapping in four models. The optimal chain strength is close to that predicted from mean energy (Fig.~\ref{fig:chainStrength}), although the dependence on chain strength is ambiguous in the CDMA case.
(right) The number of spin updates required in GD mapping (\ref{eq:GD}) from QPU samples and random samples is shown, where chain strengths are optimized based on the left plot. Most samples are local minima for 3DSG, gaining nothing by GD. For other models a small fraction of variables must be flipped to achieve a local minima. Significantly more updates (and recursion)  is required from random samples. As shown in Table \ref{table:PGS}, the local minima obtained are also less likely to achieve the target energy.}
\label{fig:pp}
\end{figure}

Results for various mapping strategies are also shown in Table \ref{table:PGS} where chain strength is optimized over a small set of values matching those plotted in Fig.~\ref{fig:pp}(left). For each instance and sample set we estimate the frequency of target states as
\begin{equation}
  p = \langle I( H(x) \leq E_T)\rangle_{x \in \mathcal{S}} \label{eq:p}\;,
\end{equation}
where I() is an indicator function $1$ for the achieved energy target, 0 otherwise. Random samples plus greedy descent (Rand+GD) is added to check the QPU samples are adding value relative to a naive algorithm, and that not all the work is being done in the post-processing. Table \ref{table:PGS} shows that mapping A is inferior to all other QPU methods, misaligned samples can be mapped efficiently to viable solutions. Table \ref{table:PGS} and other figures demonstrate R and MV behave almost indistinguishably. Rand+GD can outperform MV in some cases, but is not competitive with the QPU samples mapped by GD, particularly for the larger (and shorter chain length) models.

\begin{table}
  \caption{Median $p$ under mappings, 2283 samples in all cases}
\label{table:PGS}
\centering
\begin{tabular}{|c|c|c|c|c|}
\hline
Model & p (rand+GD) & p (GD) & p(MV) & p(A) \\ \hline
CSG(N=32) & 0.156(8) & 0.63(6) & 0.19(4) & 0.16(2) \\    \hline
CDMA(N=32) & 0.07(1) & 0.12(3) & 0.005(2) & 0.005(2) \\ \hline
BSG(N=64) & 0.008(1) & 0.38(7)  & 0.17(2) & 0.13(2) \\    \hline
3DSG(N=64) & 0.0004(4) & 0.73(5)  & 0.72(5) & 0.63(5) \\    \hline
CSG(N=64) & 0.029(6) & 0.09(4) & 0.0020(4) & 0.0013(2) \\    \hline
CDMA(N=64) & 0.019(6) & 0.027(3) & 0 & 0 \\ \hline
\end{tabular}
\end{table}


The efficiency of greedy descent and chain strength dependency is considered in Fig.~\ref{fig:pp}. Rate of target states ($p$) is a noisier statistic than energy, but they both correlate well in optimal chain strength. Energy can be an efficient proxy for maximizing $p$ where target states are rare. 
The distribution of the number of updates required to remove local excitations is also presented. QPU samples for 3DSG are already local minima at the scale N=64, whereas in other models some relaxation is typically required over a small fraction of variables. Significantly more work is required from random samples.
\begin{figure}[htbp]
  \centerline{\includegraphics[width=0.5\linewidth]{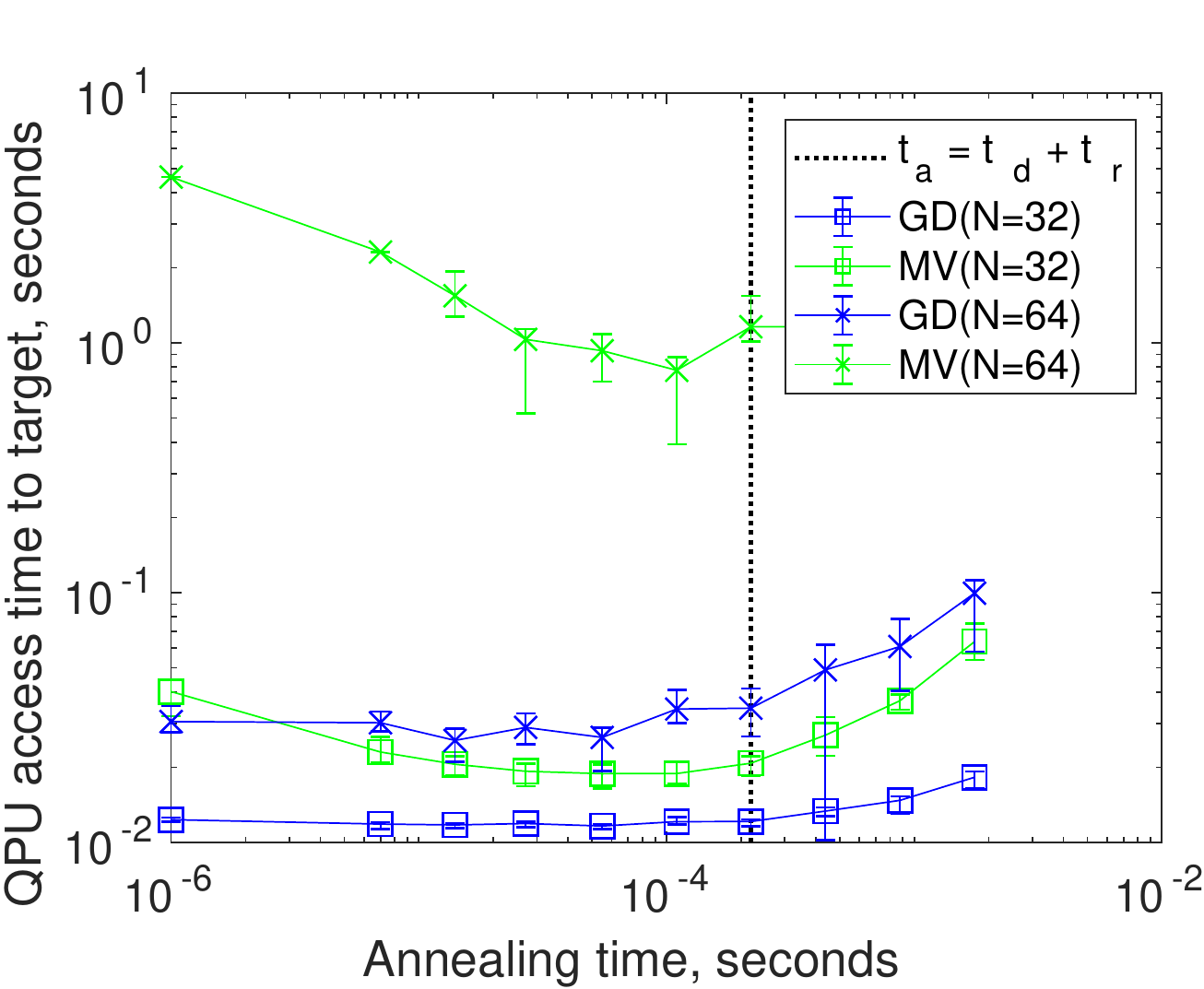}
  \includegraphics[width=0.5\linewidth]{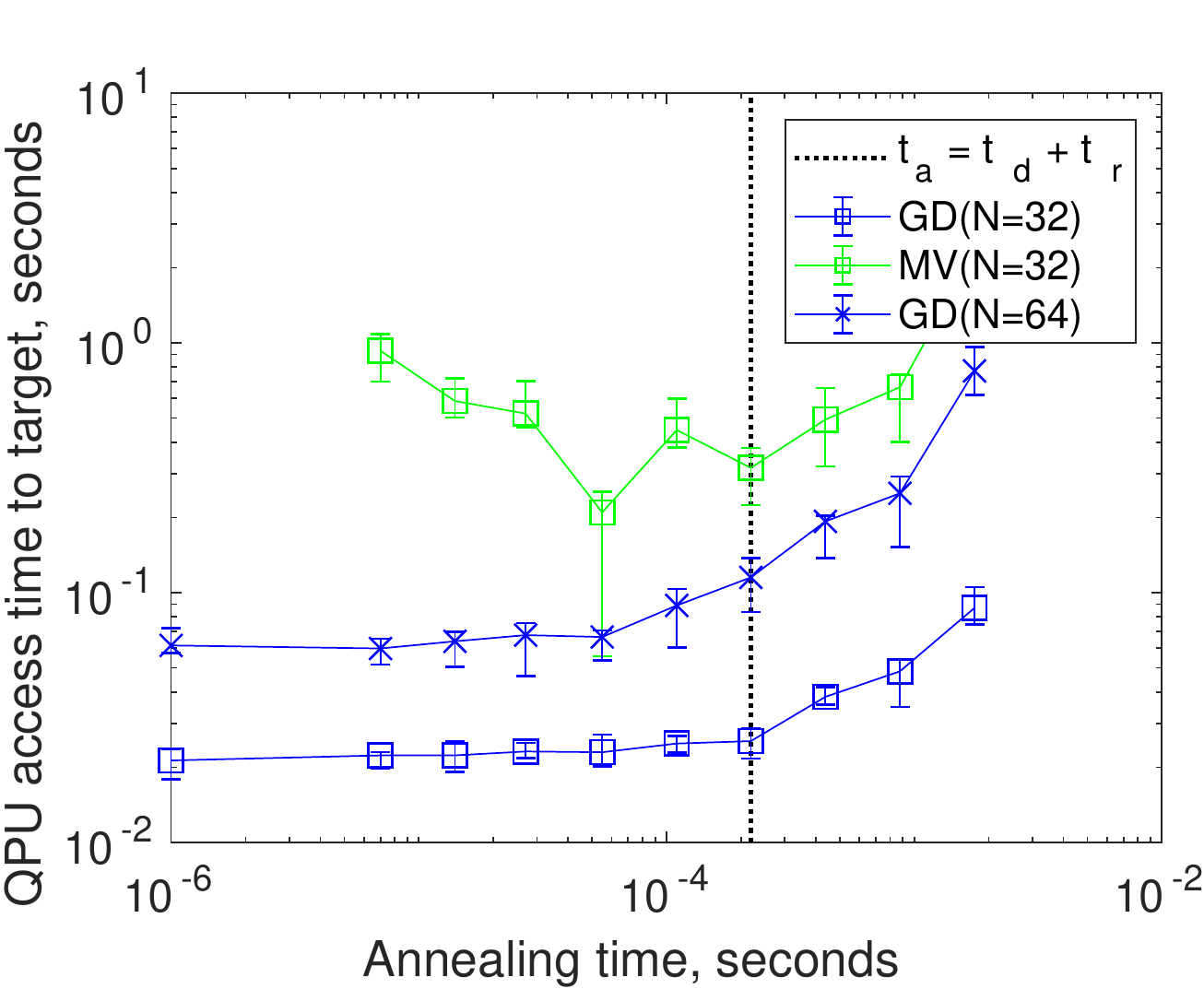}}
  \centerline{\includegraphics[width=0.5\linewidth]{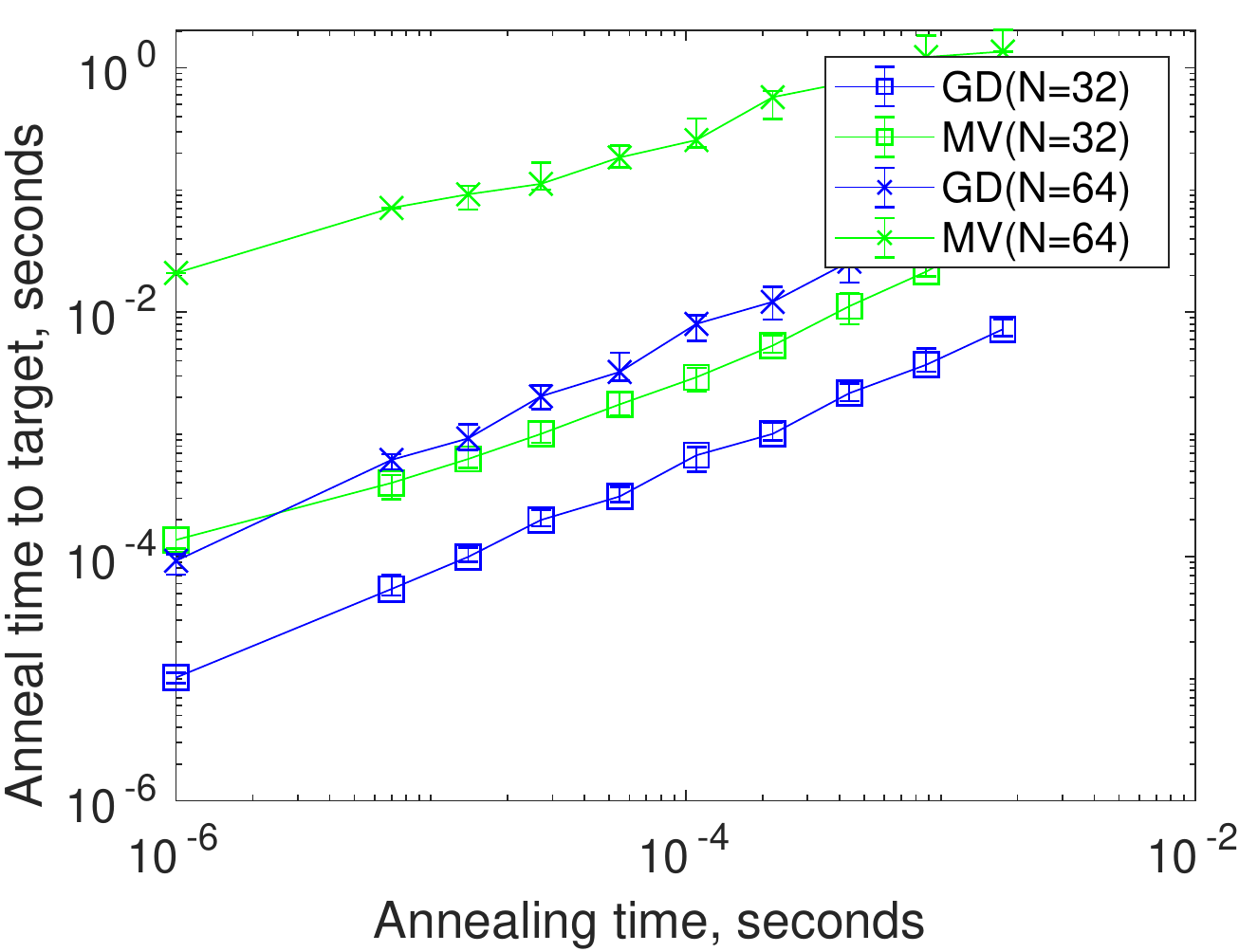}
  \includegraphics[width=0.5\linewidth]{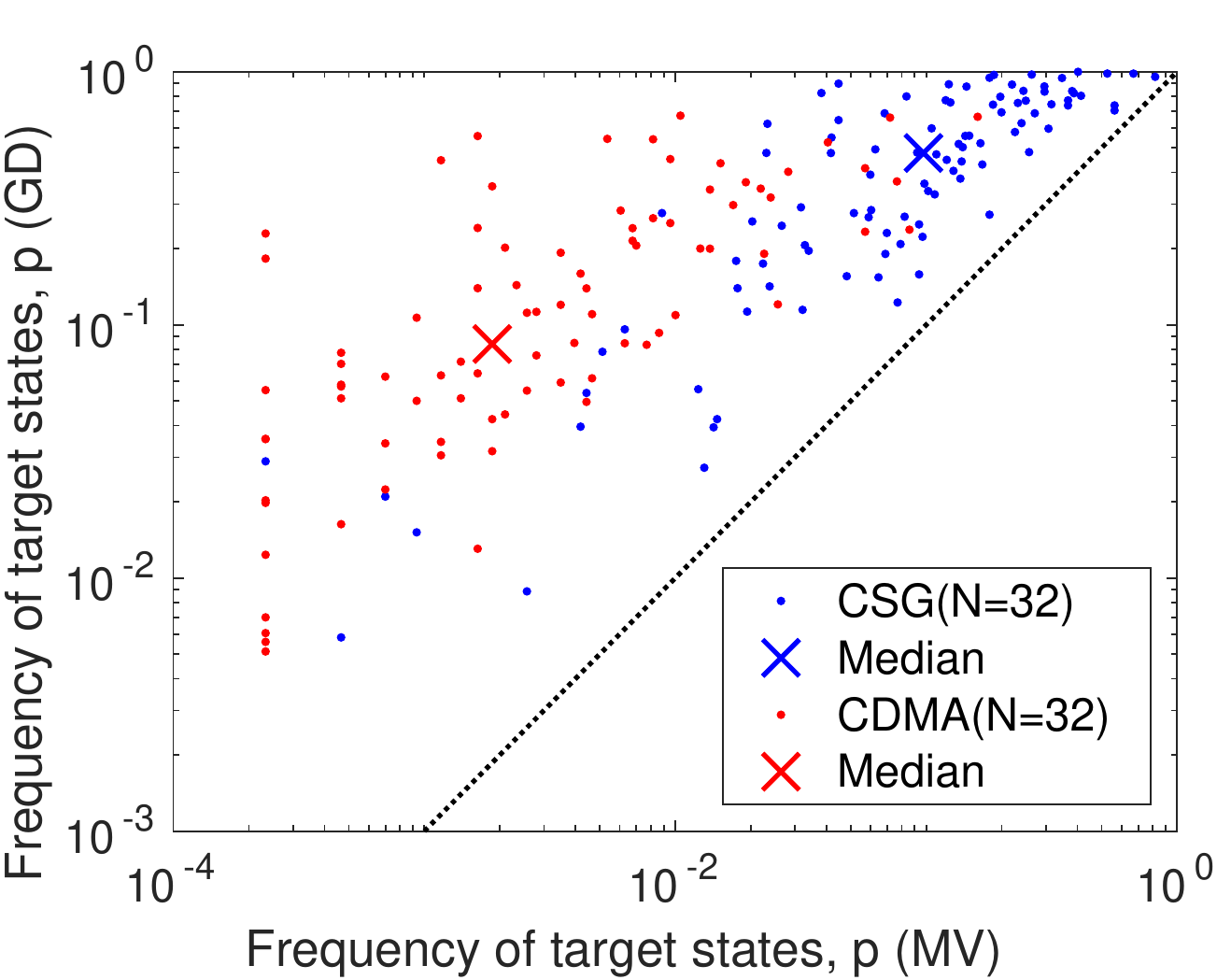}}
\caption{Considering median QPU access time we obtain optimal performance for both CSG (top-left) and CDMA (top-right) by considering $t_a \lesssim t_d + t_r$. With more powerful postprocessing (GD replacing MV) shorter anneal durations are favoured, with weaker dependence on $t_a$. (Bottom-left) Considering only annealing time-to-solution, the shortest available annealing time ($\SI{1}{\micro s}$) is optimal in median case for all ensembles (CSG shown matching top-left). (Bottom-right) At optimal QPU access time we can consider the full distribution of $p$ for MV and GD, there are strong instance to instance fluctuation.}
\label{fig:oad}
\end{figure}
In the remainder of experiments we use exclusively $\lambda=1.6\sqrt{N}$ (CSG), $\lambda=2\sqrt{N}$ (CDMA) and $\lambda=2$ (3DSG). These parameters work reasonably well across a range of experiments.

For Fig.~\ref{fig:oad} we demonstrate time-to-solution using estimates of $p$ (\ref{eq:p}) converted to a time by combination of (\ref{eq:sts}) with the QPU API timings as (\ref{eq:timings}). Ignoring all but the annealing time, the optimum is the minimum over the programmable range, the QPU access timing indicates larger values approximately inline with expectations for optimal time (\ref{eq:oad}). 
It can be seen in many cases that programming time is comparable to QPU access time-to-solution. Where overheads such as this are large, longer anneals will be preferential, so long as $p(t_a)$ is an increasing function. For all programmable sizes, in current processor technology, we can consider the non-annealing time overheads to be a significant factor in practical optimization of spin glasses and CDMA.

\section{Logical-J Compensation} 
\label{sec:ljpc}

The uniform spreading assignment (\ref{eq:us}) is the standard approach for programming given a minor-embedding. In this section we discuss a heuristic alternative based on pairwise susceptibility of logical qubits. Uniform spreading guarantees that with respect to chain aligned states, the logical energy and physical energies are aligned. However, when the transverse field is comparable to the longitudinal fields in a quantum annealing process, chain misaligned states have significant weight in the ground state wave function, and play an important role in dynamics. Qubits within a chain are not rigidly bound together, but correlated as a function of distance, and this correlation can be weak early in the anneal.

The state of the chain is encoded in the physical qubits, but peripheral physical qubits are less tightly bound to the bulk of the chain than centrally positioned ones. Since particular physical qubits mediate the coupling of chains as shown in Fig.~\ref{fig:c8clique} the transmission of chain state information (effective couplings) will depend on the topology of chain connections. Often success or failure of annealing is determined by freeze-out phenomena~\cite{quasistatic}. Biases created at these freeze-out points due to coupling topology can bias the distribution and prevent optima being found.

We can seek to compensate programmed values to make {\em effective} inter-chain couplings more homogeneous. We compensate inter-chain couplings so that effective couplings are balanced earlier in the anneal at the expense of inbalance later in the anneal (approaching $s=1$). This can be tolerated because at the end of the anneal either nothing important happens dynamically, or the dynamics are spatially local and can be post-processed away, unlike the macroscopic biases burnt in early in the anneal.

We propose a heuristic rule for the programmed values ${\hat J}_{ab} = J_{ab}\mathcal{N}/\chi_{ab}(\xi)$ parameterized by a correlation length parameter $\xi$. Pairwise-logical susceptibility is defined
\begin{equation}
  \chi_{ab}(\xi) = \prod_{i,j} \left[\frac{\sum_{i'j'}A_{i'j'}\exp\left(\frac{-|i-i'| - |j-j'|}{\xi} \right)}{\sum_{i'j'} A_{i'j'}}\right]^{\frac{1}{|C_a| |C_b|} } \label{eq:chiab}\;,
\end{equation}
where $|i-j|$ is the graph distance between two physical qubits, $|C_a|$ is the chain length, with sums and products restricted to $i,i' \in C_a$ and $j,j' \in C_b$. The normalization constant $\mathcal{N}$ can be defined as the geometric mean of $\chi_{ab}$ across all logical edges, to ensure that the typical energy of chain-aligned states is unchanged.
For the case that chains connect through only a single programmable coupler (\ref{eq:chiab}) simplifies to
\begin{equation}
  \chi_{ab}(\xi) = \sum_{i \in C_a, j \in C_b} A_{ij}\chi^i_a(\xi) \chi^j_b(\xi) \label{eq:chiab2}\;,
\end{equation}
with logical susceptibility defined
\begin{equation}
  \chi_a^i(\xi) = \prod_{j \in C_a} \exp\left(- \frac{|i - j|}{\xi |C_a|}\right) \label{eq:chia}\;.
\end{equation}

 We expect for effectively parameterized annealing that freeze-out of chains will occur at the same point that macroscopic features of the solution space are determined~\cite{Venturelli2015a,quasistatic}. Therefore, the chain length parameter is expected to be close to the chain length $\xi \approx |C_a|$ for use of this heuristic. 

Assuming a quasi-static model of the annealing process, $\chi_a^i(\xi)$ can be interpreted as the susceptibility of the chain to a field applied at site $i$ at the freeze-out point ($s^*$), the final point in the anneal with fast dynamics\cite{quasistatic}. Consider that all physical qubits in the chain are equally representative of the state of the chain. The response of a chain to a change at physical qubit $i$ can be considered the average of the responses of all the physical qubits comprising the chain. The response of physical qubit $j$, to a perturbation on qubit $i$, is---by the fluctuation-dissipation theorem---equal to the correlation. Considering either a 1D quantum (or classical) model we anticipate the correlation to decay approximately exponentially according to the distance between the two physical qubits $C_{ij} \propto \exp(- \frac{|i - j|}{\xi})$. Under this sequence of steps, we have the interpretation of (\ref{eq:chia}) as the logical qubit susceptibility, defined as a geometric mean of the susceptibility on each qubit compromising the chain.

Pairwise-logical susceptibility extends this single chain notion. Consider that we have a fluctuation in the state of some physical qubit $j'$ on chain $1$ and wish to know the response of state $i'$ on a second chain, where the two chains are coupled by $J_{ij}$ (see Fig.~\ref{fig:c8clique}). The response between chains is mediated by $J_{ij}$, but within each chain by $C_{jj'}$ and $C_{ii'}$. Taking a geometric mean gives~(\ref{eq:chiab}). In the case of multiple couplings between two chains, we take a mean over the paths (appropriate for a weak coupling limit).

Pairwise-logical susceptibility is a measure of how much weaker the coupling is between chains relative to what would be expected for a pair of directly coupled physical qubits. We achieve our aim of homogenization of the coupling strength across couplers of various topology by scaling the programmed values inversely to this value.

More detailed derivations and generalizations of this method are discussed in Appendix \ref{sec:susccompensation}. This method is closely connected to the spectral compensation method used to compensate interactions in a 3DSG study\cite{Harris162}. As shown in Appendix \ref{sec:spectralcompensation}, the results are compatible in most cases. 

\subsection{Results}

\begin{figure}[htbp]
  \centerline{\includegraphics[width=0.5\linewidth]{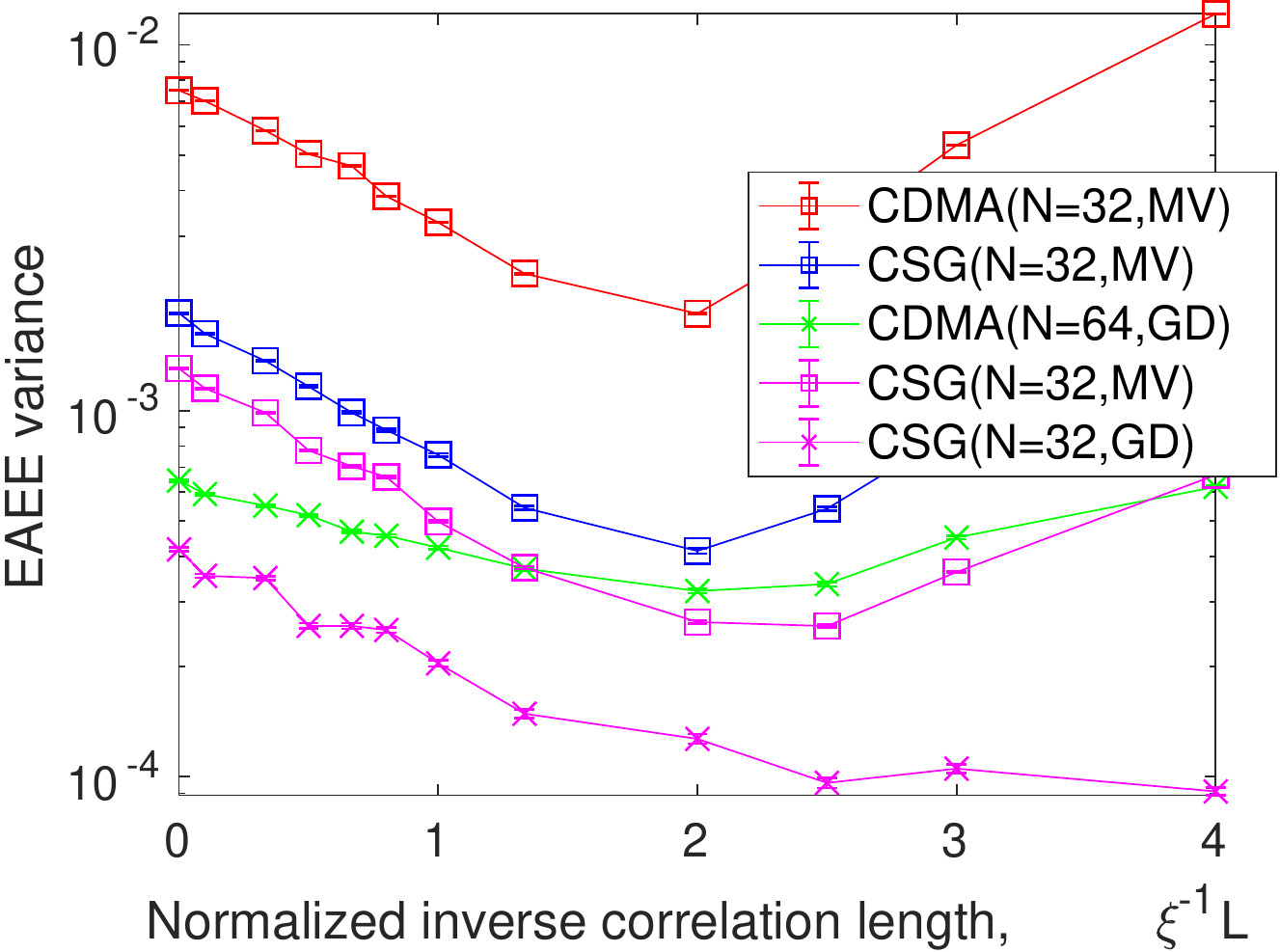}\includegraphics[width=0.5\linewidth]{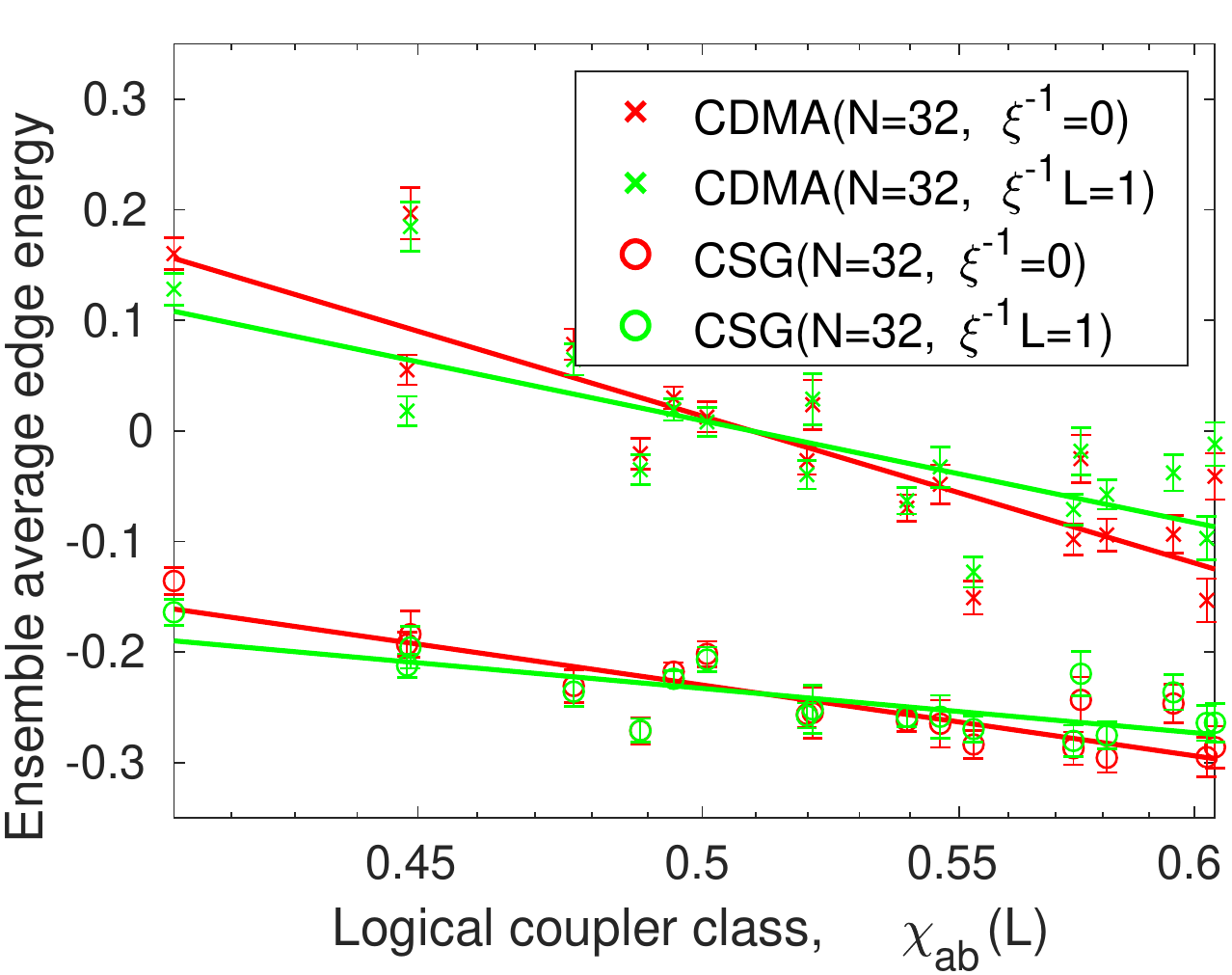}}
  \centerline{\includegraphics[width=0.5\linewidth]{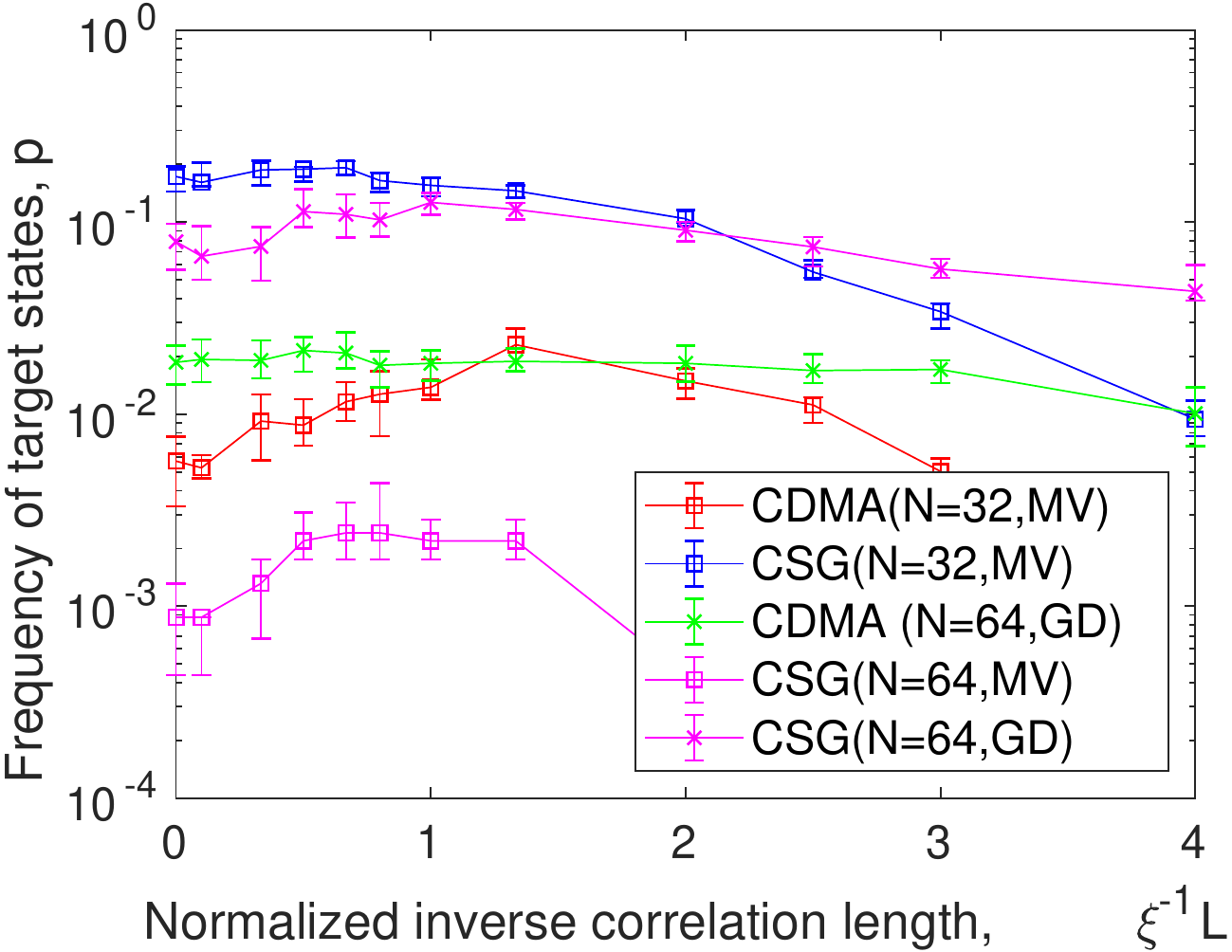}\includegraphics[width=0.5\linewidth]{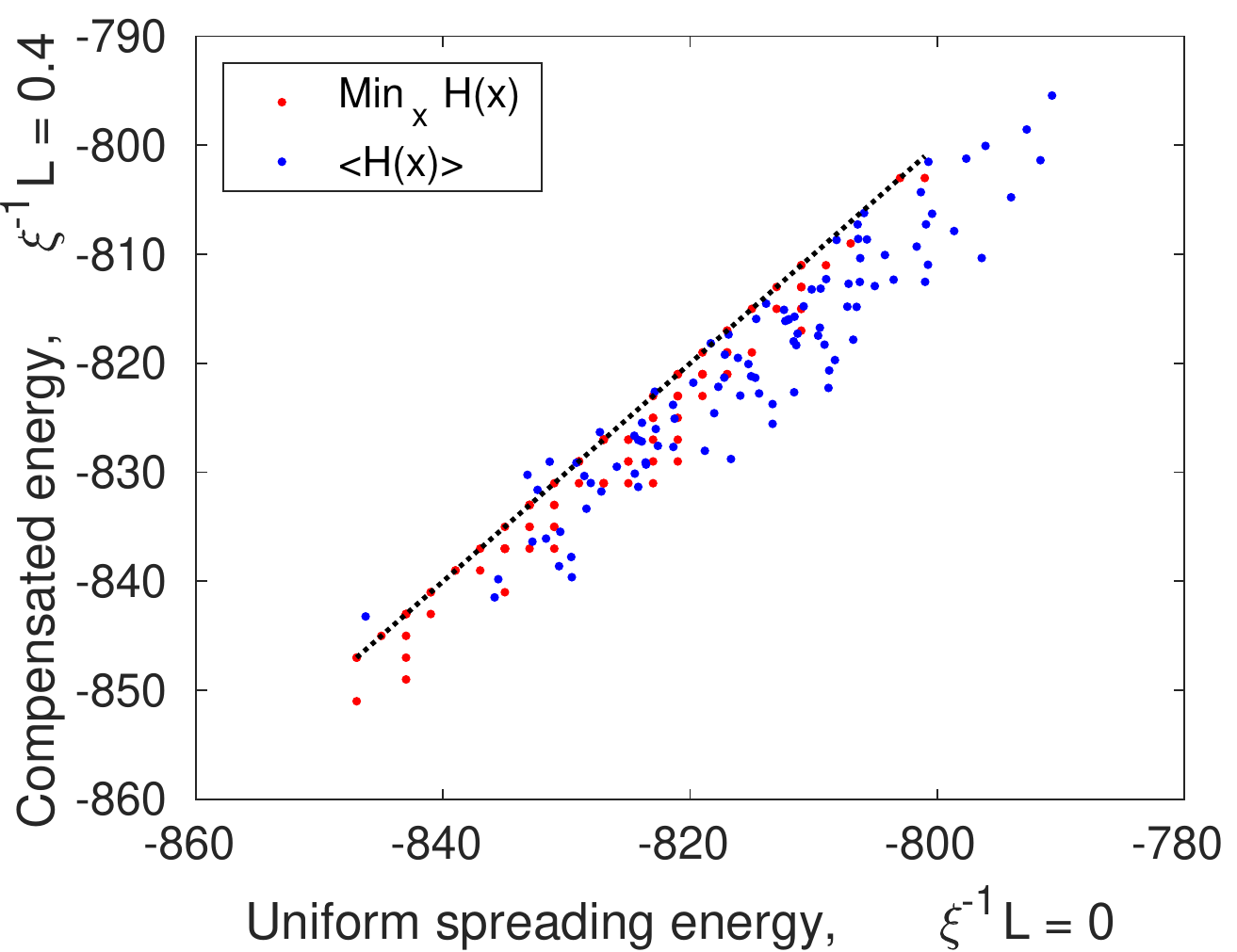}}
\caption{(Top-left) By symmetry, variance of $e_{ab}$ should approach zero when logical-Js are functioning ideally. A substantial imbalance exists for $1/\xi = 0$ (uniform spreading) which is corrected by compensation. (Top-right) We can break down the contributions as a function of topology, correlations that are connected through peripheral qubits are significantly more likely to be frustrated (larger $e_{ab}$). After correction we approach the desired result (0 for CDMA), uniform for CSG, whilst also decreasing energy. (Bottom-left) Probability to reach the target energy is increased as inhomogeneity is corrected, although the optimal $\xi$ value is smaller than that considering only variance of $e_{ab}$. (Bottom-right) The case of 3DSK at full lattice scale ($8\times 8 \times 8$) is demonstrated on a per instance basis. Considering the energy distribution before and after compensation, there is significant improvement, the min energy achieved is at best equal to the compensated case in all instances. Strong performance in the cubic lattice is supported by related studies~\cite{Harris162}.}
\label{fig:correlationBalancing}
\end{figure}
As in previous sections we study the median properties with respect to 100 instances, using $t_a=t_d+t_r$ and for each instance drawing sample sets of size $\lfloor 1/(t_a +t_d+t_r)\rfloor=2283$ samples.

As a simple demonstration of the bias attributable to uniform spreading, and the restoration of homogeneity endowed by compensation, we will consider the ensemble average edge energy (EAEE), defined per logical edge as
\begin{equation}
  e_{ab} = \overline{J_{ab} \langle x_i x_j \rangle_{x \in \mathcal S}} \label{eq:EAEE}
\end{equation}
where ${\overline \cdot}$ denotes an instance average, and $\langle \cdot \rangle$ denotes the usual sample average. 

Considering CDMA, CSG, and BSG models on the logical problem (1 physical qubit per logical qubit) there are strong fluctuations instance to instance, but after averaging over instances all edges are indistinguishable and contribute evenly to the ensemble average energy. $e_{ab}$ is constant---this is true throughout the anneal even in the case of decoherence or thermalization, provided these phenomena do not break spatial symmetry. For the CSG and BSG models the constant is precisely $\overline{\langle H(x) \rangle}/[N(N-1)/2]$. For the CDMA problem, by contrast, if our optimization works well we find solutions of the form $x\approx b$, in which case $\overline{J_{ij} \langle s_i s_j \rangle} = 0$. However, when we repeatedly use the same minor-embedding for these problems we expect to see a symmetry breaking following our susceptibility arguments.

A simple measure of the symmetry breaking is given by the variance of $e_{ab}$, and success of the compensation scheme can be measured by the deviation of this quantity from $0$. However, when it comes to a heuristic of this kind we should be careful that balance is not restored at the expensive of energy or time-to-solution. It is easy to achieve zero variance by fully randomizing the samples and losing all information on the low-energy states. Therefore in tandem we can consider energy. We can work at near optimal parameters for the uniform spreading heuristic ($\xi=\infty$), and demonstrate improvements as compensation is switched on.

In Fig.~\ref{fig:correlationBalancing} we show the variance of $e_{ab}$ and energy under compensation for the case of CDMA and CSG. Variance is reduced using GD, but a signal remains, and both homogeneity and energy are improved through compensation.

We can also consider how effective the compensation is per coupling pattern. We can classify each coupling between logical qubits by the induced subgraph considering only the two chains and the couplings between them. Two such subgraphs are in the same class if they are automorphic. The number of such patterns varies widely according to the graph-minor topology as shown in Table \ref{table:cct}.
\begin{table}
  \centering
  \caption{Chain connectivity topologies}
  \label{table:cct}
\begin{tabular}{|c|c|c|c|c|}
\hline
Topology & Variables,N & Chain length,L & Logical coupling patterns\\ \hline
Clique & 32 & 9 & 18 \\ \hline
Clique & 64 & 17 & 51 \\ \hline
Biclique & 64 & 8 & 10 \\ \hline
3D cubic & 64 & 4 & 3 \\    \hline
3D cubic & 512 & 4 & 3 \\    \hline
\end{tabular}
\end{table}

We can take an average over couplers in a given automorphic class, and classify each class according to $\chi_{ab}(1)$ for some element $ab$ in the class. The result is shown in Fig.~\ref{fig:correlationBalancing} (top-right). Frustration is indeed disproportionately allocated over couplers that connect through extremal points (small $\chi_{ab}$), we see that applying an appropriate compensation of $1/\chi$ between one and two makes coupler performance more homogeneous.

In Fig.~\ref{fig:correlationBalancing} (bottom-right) we show the mean energies and best minimal achieved for all instances of the 3DSG problem ($8\times 8 \times 8$) with and without logical-J compensation. We allow for a small number of edge and variable vacancies (in line with defect patterns on the chip). Energies are significantly improved not only in the median but in almost all instances.



\section{Conclusion}
\label{sec:conclusion}

In this paper we have demonstrated that inhomogeneities in the patterns of chain couplings produce biases in the sampling distribution, and that compensation by a simple heuristic can reduce biases and improve optimization performance. Using a simple heuristic scheme based on the assumption that the distribution reflects a freeze-out point in the anneal where chains are rigid (but not fully correlated) allows results to be improved in hard random problems: spin glasses of various topology, and CDMA. The rigidity of the chain is described by the correlation length, and a value close to the chain length produced good outcomes. We simultaneously demonstrated the impact of chain strength, anneal duration, and physical-to-variable mappings on reaching low-energy states in these models. A simple GD approach (\ref{eq:GD}) was found to be important, and a chain strength tuning rule (\ref{eq:chainStrength}) was found to work well across a range of problem types and sizes. We argued that one might expect an anneal time comparable to the sum of read-out and delay times would be optimal considering QPU access time (\ref{eq:oad}), but in most cases found optimal values that were slightly shorter.  

The performance of cliques at full scale is a stringent test of quantum annealing with minor-embedding. To embed $N=64$ cliques on a Chimera-structured graph requires chains of length $17$, and requires these to approximate a spin state. Forthcoming generations of D-Wave quantum computers are also constrained in connectivity, although chain lengths for cliques and other topologies are significantly shorter due to technological advances~\cite{Boothby:19}. Nevertheless, in this paper we have shown that CDMA and CSG models are solvable at the maximum programmable scale in D-Wave 2000Q computers.
We have not compared against state of the art competition, but we can compare the performance to earlier QPUs where the largest clique studied was $N=30$, and with a runtime to 99\% confidence at this size of {\em at best} \SI{1}{\milli s} (\cite{Venturelli2015a}, Fig.~3). Setting chain strength to $\lambda_0=1.6$ and using MV (as done in that earlier study) we can obtain a result at size $N=32$ of $\sim\SI{100}{\micro s}$ (Fig.~\ref{fig:oad} bottom-left). This improvement leverages QPU improvements such as faster anneals and extended {\em J}-range. Further significant advantages can be gleaned replacing MV by GD (Fig. \ref{fig:oad}), with logical-J compensation (Fig. \ref{fig:correlationBalancing}), or optimization of chain strength. It should also be noted that at a scale $N=64$ we can solve typical case of CSG within \SI{20}{\milli s} of annealing time, in strong contrast to scaling predictions from the earlier studies.

The form of logical-J compensation (\ref{eq:chiab}) is derived based on susceptibility arguments, and has the strengths of a simple phenomenological interpretation. For 1D chains we expect an exponential model to be a reasonable for both quantum and classical processes, and robust to noise. The exponent form can be derived analytically in special case, and generalizes to other chain shapes, as shown in Appendix \ref{sec:susccompensation}. It also correlates very well with a spectral compensation method discussed in Appendix \ref{sec:spectralcompensation}. However, derivation relies on a 2-chain level approximation and the notion that effective coupling is determined by a single quasi-static freeze-out point for all edges, whereas the situation is typically more complicated. The approximation is demonstrated here for random problems that are relatively homogeneous with identical (equal chain length and shape) logical qubits. Although the intuitions provided by a susceptibility framework should remain valid, verification is necessary in other contexts.

We have chosen CDMA and spin glasses as exemplars because they are well studied canonical random problems, with well understood equilibrium and dynamical scaling properties in $N$, both for quantum and classical cases~\cite{PhysRevE.90.012126,TanakaCDMA,Nishimori2001,qpt}. The prediction for large $N$ is that these problems are hard for a variety of heuristics owing to the nature of the phase transitions. These transitions apply in typical cases of the problem and so it is sufficient to look at median performance to access relatively hard problems. However, it should be noted that the cases studied herein are relatively small, and do permit efficient solution finding by classical heuristics. Furthermore, there is quite large variability in statistics, as indicated in, for example, Figure \ref{fig:oad}. With regards tuning of parameters there is the usual caveats: we tune here for the median, which means we may solve outliers poorly. To solve a plurality of instances one has to tune for the harder instances: this can mean longer anneal duration, and modified chain strengths, even if we can be confident the variability will become smaller for large $N$. 

The D-Wave 2000Q, and embedding procedures, have a large number of protocol parameters that can be tuned. These include recent feature advances that allow manipulation of the anneal schedule shape, manipulation of the schedule for biases relative to couplers, delays or advances over the schedule for individual qubits, and even reverse annealing allowing preparation of a classical initial condition. The annealing protocol can be tuned within reason to enhance performance~\cite{King2016,10.1145/3341302.3342072}. In this paper we vary only of a handful of parameters for which we can provide fairly generic advice. 
For random problems one might in principle tune a modest number of parameters on a test set, and then apply them across a variety of new instances. This can allow an ensemble-level fine tuning of quantities like chain strength. We have focused on performance over single programmings, without catering to the specific structure of any particular instance, or allowing the possibility of refined tuning through sampling feedback.

In the case of quantum annealing, analog error sources mean that biases can exist in the Hamiltonian. Multi-programming mitigation strategies can be employed to average over these: spin-reversal transforms and the use of multiple embeddings are sensible strategies. The use of multiple embeddings, or automorphisms in the case of cliques bicliques and cubic lattices, not only averages over noise in the processor, but averages over the logical-J inhomogeneities that have been identified as problematic. These transformations do not rely on measurements of the sampling distribution, but there are also innumerable ways to incorporate data-driven tuning approaches across several programming cycles.

We have shown that CDMA and a variety of spin-glass problems can be solved on the QPUs. To do so efficiently requires choices to be made in chain strength, anneal duration, physical-to-logical mapping, and minor-embedding, and we have provided rules for these choices. We showed that one consequence of minor-embedding is inhomogeneity in the coupling strengths. The inhomogeneity is a function of the connectivity between chains, and we demonstrated a rule to compensate for this that improves performance in optimization. In the case of cliques, we can solve instances at the largest size in a single programming, and demonstrated progress with respect to similar experiments on earlier QPU generations.



\bibliographystyle{IEEEtran}
\bibliography{chainCompensation3}%
\appendix


\subsection{Logical-J Compensation, Spectral Method}
\label{sec:spectralcompensation}

\begin{figure}[htbp]
\centerline{\includegraphics[width=0.7\linewidth]{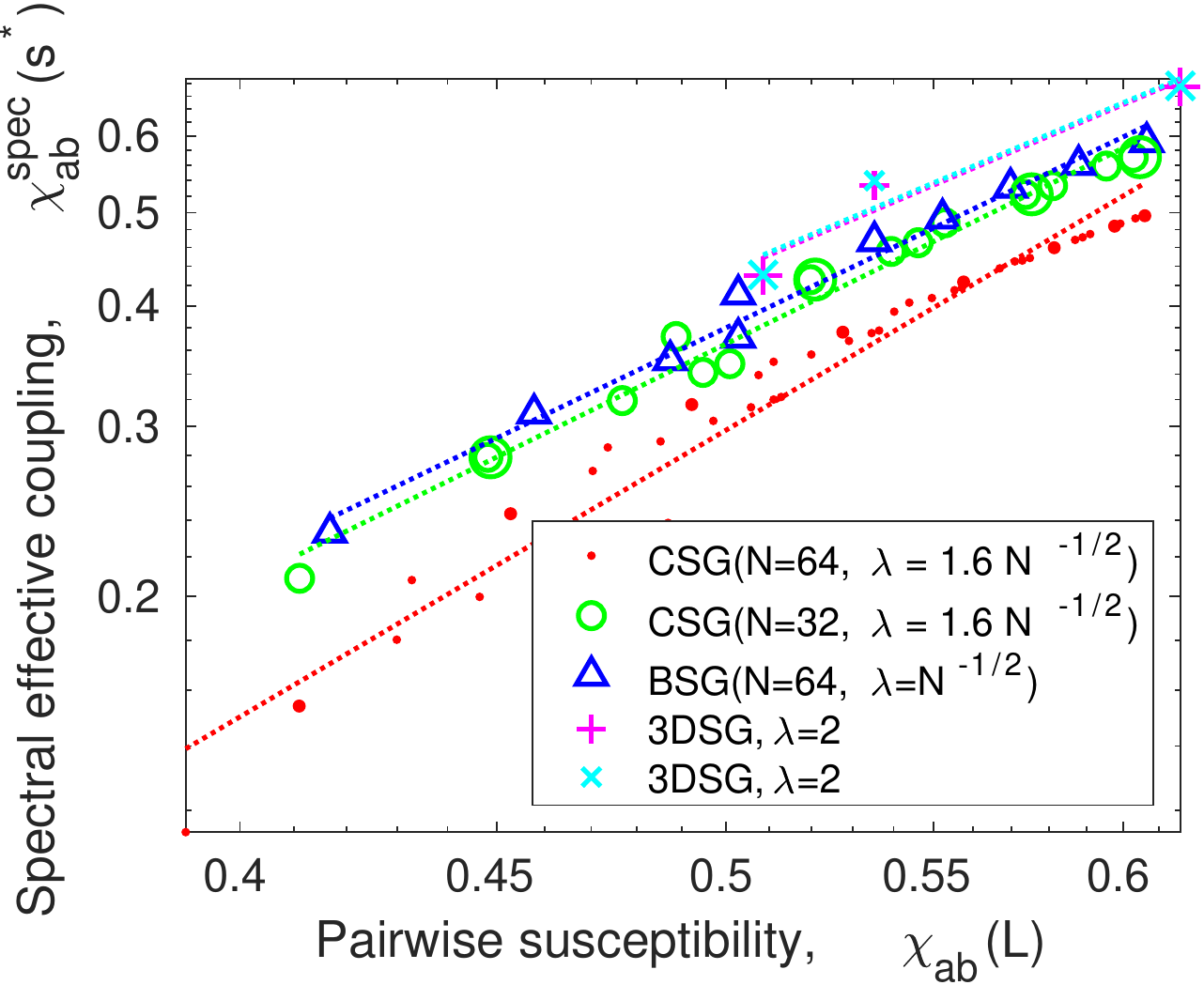}}
\caption{Many problems freeze-out at $A(s^*) \approx B(s^*)R\lambda$, when chain coupling strength approximately matches the transverse field, the phase transition point for long chains. We scatter the effective coupling strength against the susceptibility compensation form at $\xi=L$. Larger markers indicate the minority of doubly connected chains, which behave similarly to the singly connected chains. The behaviour is strongly correlated, but spectral compensation deviates from the simple exponent description (as indicated [approximately] by deviation from straight line fits). In 3DSG we demonstrate two values one tuned for optimization ($\lambda=2$, exploiting maximum energy scales), and another tuned larger for better logical emulation ($\lambda=2/0.3$)~\cite{Harris162}, in each case compensation is comparable demonstrating approximate linearity of the effective coupling strength on the programmed value.}
\label{fig:specComp}
\end{figure}

We briefly summarize the spectral method for chain compensation outlined in \cite{Harris162}. Whereas in that study the spectra of an rf-SQUID model was studied, herein attention is restricted to an Ising model approximation. Consider a single qubit problem mid anneal, with zero longitudinal field and isolated from interactions with other qubits. This problem is defined $H_i(s) = A(s)\sigma^x_i$, and has a gap between ground $E_0$ and first excited $E_1$ energies defined as $2A(s)=E_1-E_0$. Consider also an edge in a problem without an embedding (1 physical qubit per variable) at zero longitudinal field and isolated from other interactions:
\begin{equation}
  H_{ij}(s) = B(s) J \sigma^z_1 \sigma_2^z  + A(s)[\sigma^x_1 + \sigma^x_2] \;.
\end{equation}
The gap between the first and second excited energies defines the coupling strength $2 B(s) J = E_2-E_1$. 
Now consider an isolated logical qubit
\begin{equation}
  {\hat H}_a = A(s) \sum_{i \in C_a} \sigma^x_i + B(s) \lambda \sum_{i<j \in C_a} A_{ij} \sigma^z_i \sigma^z_j\;.\label{eq:1chain}
\end{equation}
We can identify the first gap of this system as $2 A_{eff}$, by analogy with physical qubit states.
Similarly a system of two logical qubits can be written
\begin{multline}
  {\hat H}_{ab}(s) =  A(s)\sum_{x = \{a,b\}} \sum_{i \in C_x }\sigma^x_i + \\B(s) \left[{\hat J}_{ab}\!\!\!\sum_{i \in C_a,j \in C_a} A_{ij} \sigma^z_i \sigma^z_j + \lambda \!\!\!\sum_{x = \{a,b\}}\sum_{i<j \in C_x} A_{ij} \sigma^z_i \sigma^z_j\right]\;. \label{eq:2chain}
\end{multline}
Again we can make an identification between the lowest eigenstates of this system and those of the physical-qubit pair thus $2 B(s) J^{eff}_{ab}(s) = E_2-E_1$.
Note that the spectrum here is a function of the pattern of connectivity between the logical qubits.

Choosing chain strength large enough we can separate these lowest-energy states from higher excitations, and expect them to dominate dynamics in an adiabatic annealing framework. 

The spectral effective coupling is reduced throughout the anneal by a factor 
\begin{equation}
  \chi_{ab}^{spec}(s) = J_{ab}^{eff}(s)/J_{ab}\;. \label{eq:chispec}
\end{equation}
Supposing at some point in the anneal ($s=s^*$) dynamics become slow then the distribution reflects coupling strengths at this point. 
Assuming a linearized dependence of effective coupling on the programmed coupling, we can modify the couplings replacing (\ref{eq:chiab}) by (\ref{eq:chispec}). 
An iterative method can be used to refine this in cases of a significant non-linearity, this rarely proves necessary in well-tuned annealing protocols. 

The spectral compensation (\ref{eq:chispec}) is strongly correlated with the susceptibility compensation (\ref{eq:chiab}) as shown in Fig.~\ref{fig:specComp}, but does differ from a simple exponent description. After appropriate tuning of the freeze-out point ($s$) and chain strengths results are very similar in the models studied replacing susceptibility compensation by spectral compensation. Each method has strengths and weaknesses, and we focus only on susceptibility compensation in this paper. Three notable downsides of a spectral approach are (1) the argument does not straightforwardly generalize for non-zero $h$, (2) the argument is tied to properties of coherent ground (and lowest level) states only, (3) the compensation required for the weakest chain couplings is significantly larger than for the simple exponent case (and large compensations risk non-linearities and stretch the programming energy budget). However, it should be clear that both methods allow space to develop towards one another, and qualitatively capture the same idea. For example, we could replace our simple exponent susceptibility with an exact susceptibility of ground state wave functions as a hybrid; many other possibilities exist.

\subsection{Logical-J Compensation, Susceptibility Method}
\label{sec:susccompensation}

In this appendix we describe in greater technical detail the principles underlying the logical-J compensation heuristic (\ref{eq:chiab}). We begin by deriving this formula assuming a classical model parameterized by inverse temperature $\beta$.

Consider single and pair logical qubit models, as per Appendix \ref{sec:spectralcompensation}. We can consider susceptibility of these objects directly, but for simplicity consider a classical approximation. Taking $A(s) = 0$ and $B(s)\lambda = 1$ and introducing an inverse temperature parameter $\beta$ we can approximate the distribution of physical states on an uncoupled chain by
\begin{equation}
  P_a(z) \propto \exp(\beta \lambda \sum_{i,j \in C_a} z_i z_j) \;.
\end{equation}
If the pattern of connectivity among qubits is tree like, a special case being a 1D sequence (chain), we can define the correlations on the chain explicitly as $C_{ij} = \tanh(\beta \lambda)^{|i-j|}$.  $\chi_a^{i}(\xi)$ is defined as a geometric mean of physical qubit correlations, provided we identify $\xi^{-1} = -\log(\tanh(\beta\lambda))$, we arrive at our first expression (\ref{eq:chia}).

We can consider in the same vein a pair of logical qubits with classical fluctuations
\begin{equation}
  P_a(z) \propto \exp(-\beta  {\hat J}_{a,b}\sum_{i \in C_a, j \in C_b} A_{i,j} x_i x_j)\prod_{x=a,b}P_x(z)\;.
\end{equation}
Given this model we could again explicitly constructed $\chi_{ab}$ as the geometric mean of correlations between all pairs of qubits on each chain. Supposing there is a unique path between any pair of qubits ($i'$ and $j'$) mediated by a coupling between $i$ and $j$ ($i,i' \in C_a, j,j' \in C_b$), we find
\begin{equation}
  C_{i',j'} = \tanh(\beta {\hat J}_{ab}) \tanh(\beta \lambda)^{|i-i'|}\tanh(\beta \lambda)^{|j-j'|}\;.
\end{equation}
The correlation between qubit states by this path can be taken as a geometric mean of the paths between physical qubits
\begin{equation}
  C_{a,b} = \tanh(\beta {\hat J}_{ab}) \prod_{i' \in C_a}\tanh(\beta \lambda)^{\frac{|i-i'|}{|C_a|}}\prod_{j' \in C_b}\tanh(\beta \lambda)^{\frac{|j-j'|}{|C_b|}}\;.
\end{equation}
Given a direct coupling (without embedding) we would anticipate $C_{a,b} = \tanh(\beta {\hat J}^{eff}_{ab})$, so this equation provides an interpretation for $J_{eff}$. Linearizing in ${\hat J}$, and identifying the latter two terms as $\chi^i_a \chi^i_b$ with correlation length derived from $\beta$ as per the single logical qubit case we have (\ref{eq:chiab2}).

Beyond chains connected by a single coupler, we must consider for our studies chains connected by a pairs of couplers. In this case a correlation $C_{ab}$ is mediated by multiple paths.
To accommodate this in a simple way we can suppose the coupling is weak, so that we can linearize in the coupling strength between chains. Each correlation between qubits is then a sum of paths, leading to expression (\ref{eq:chiab}). 

This entire calculation can proceed with arbitrarily structured logical qubits (beyond chains), arbitrary complicated patterns of connectivity, and typically without linearization or other approximations: We can assume an effective temperature,\footnote{The effective temperature parameter needn't be tied to a correlation length or physical temperature interpretation. It can be tuned to approximate the impact of quantum, thermal and noise sources, in so far as they impact correlations. } and from this explicitly construct and average over correlations between physical qubits in different chains. Provided the logical pairs are not too large (more importantly, provided the graph width is not too large) the calculations are efficient. A generalization might also consider a quantum Boltzmann distribution, involving both thermal and quantum fluctuations, from which correlations can also be calculated (though less efficiently for larger logical qubits). In this case the parameterization might be interpreted in terms of a quasi-static freeze-out at some pertinent point in the schedule $s$: $\{A(s)/k_B T,B(s)/k_B T\}$. 

At several points in the derivation we make use of geometric means. One place is in the compensation normalization $\mathcal{N}$. If $h=0$, as in the spin-glass models, then the normalization is made redundant by the chain strength parameter. However at non-zero $h$ as in CDMA, the choice determines the scale of $h$ relative to $J$. We choose the geometric mean here, and in other places, in order to capture a typical value for the quantities being averaged. The quantities vary exponentially, and so a linear mean may be dominated by only a small number of outliers. The geometric mean captures a bulk behaviour and was found to give a better fit to empirical data patterns for the effective coupling.

We have derived our results by arguing that all physical qubits on the chain might be considered equally representative of the logical state, but this is not essential to the arguments in many cases. Some physical qubits may be more computational relevant than others - also with impacts for the mappings.

The insight that chains are not rigid at mid-anneal has consequences beyond logical-J compensation, and other deviations from the uniform spreading approach (\ref{eq:us}) are worth exploration. These might include variation of the chain strength or variation of ${\hat h}$, and consideration of susceptibility alongside single or pairwise qubit models can be a means to do this. 
Consider that $\chi^i_a$ (\ref{eq:chia}) is introduced to describe the response of the chain $a$ to a perturbation at $i$.
The source of this fluctuation at $i$ could be considered as the external field $h\sigma^z_i$. In uniform spreading (\ref{eq:us}) we divide $h$ equally among the physical qubits of the chain, and we can consider the response of the entire chain to this choice given by $h_{eff} = h \chi_a$; where $\chi_a= [\prod_{i}\chi_a^i(\xi)]^{1/|C_a|}$. Thus as well as an effective coupling we have an effective external field. Note that this scaling factor is the same for all chains (if they are the same length). Nevertheless, compensating so that effective $h$ and $J$ are better balanced is a possibility. Time-dependent gain in Hamiltonian biases is a feature in the D-Wave API that could be used for this, allowing balancing throughout the entire anneal (rather than at a single point). One can also use this susceptibility framework to rebalance the external field across the physical qubits, so that the weaker response at the chain periphery is mitigated.

We have argued that the correlation length parameter should be approximately equal to chain length whenever the annealing protocol is properly tuned. In computationally challenging problems, arguments have been made that chains freeze at the same point that the problem becomes macroscopically ordered~\cite{Venturelli2015a,quasistatic}. Consider for example the CSG model, minor embedded on Chimera graphs, as in this paper. If chains are not rigid as macroscopic order emerges, the problem will break up into two dimensional domains breaking chains across the boundaries. Similarly, if chains freeze-out early then problem energy terms are not felt, and a high energy logical state will be obtained. The only way to access this high performance regime (if it exists at all) is to set correlation length approximately equal to chain length. One of the strengths of the logical-J compensation proposal is that little problem specific tuning is required in this regard.

\end{document}